\pdfoutput=1  % serve per imporgli di andare in pdflatex, utile quando ci sono figure grosse in .eps, e arXiv chiede di convertirle in .pdf per farlo %compilare in pdflatex
\def\Pb{P_{\rm b}}

\def\rfr#1{Equation~(\ref{#1})}
\def\rfrs#1#2{Equations~(\ref{#1})~to~(\ref{#2})}
\def\Rfr#1{Eq. (\ref{#1})}

\def\virg#1{``#1"}

\def\eqi{\begin{equation}}
\def\eqf{\end{equation}}
\def\eqia{\begin{eqnarray}}
\def\eqfa{\end{eqnarray}}

\def\lb#1{\label{#1}}
\def\kap{\bds{\hat{S}}}

\def\bds#1{\mathbf{#1}}
\def\ton#1{\left(#1\right)}
\def\qua#1{\left[#1\right]}
\def\grf#1{\left\{#1\right\}}

\documentclass[onecolumn]{aastex}
\usepackage{morefloats}
\usepackage[title]{appendix}
\usepackage{hyperref}
\usepackage{booktabs}
\usepackage[table,xcdraw]{xcolor}
\usepackage{multirow}
\usepackage{rotating,tabularx}
\usepackage{float}
\usepackage{enumerate}
\usepackage{rotating}
\usepackage[polutonikogreek,english]{babel}
\usepackage{amsmath,starfont,textgreek,w-greek,wasysym}
\usepackage{amsthm}
\usepackage{amscd,lineno}
\usepackage{amssymb,dsfont}
\usepackage{graphicx,graphics,epsfig}
\usepackage{txfonts}
\usepackage{xr-hyper}
\RequirePackage{color}

\newcommand{\emaila}{lorenzo.iorio@libero.it}

\allowdisplaybreaks[1]

\begin{document}

\title{Classical and general relativistic post-Keplerian effects in binary pulsars hosting fast rotating main sequence stars}

\shortauthors{L. Iorio}

\author{Lorenzo Iorio\altaffilmark{1}}
\affil{Ministero dell'Istruzione, dell'Universit\`{a} e della Ricerca
(M.I.U.R.)-Istruzione
\\ Permanent address for correspondence: Viale Unit\`{a} di Italia 68, 70125, Bari (BA),Italy}
\email{\emaila}

\author{Michel Rieutord\altaffilmark{2} }
\affil{IRAP, Universit\'e de Toulouse, CNRS, UPS, CNES,
14, avenue Edouard Belin, F-31400 Toulouse, France}
\email{Michel.Rieutord@irap.omp.eu}

\author{Jean-Pierre Rozelot \altaffilmark{3} }
\affil{Universit\'e C\^ote d'Azur, Observatoire de la C\^ote d'Azur,  CNRS, Nice $\&$ 77, Chemin des Basses Mouli\`{e}res 06130 Grasse, France}
\email{Jean-Pierre.ROZELOT@univ-cotedazur.fr}

\author{Armando Domiciano de Souza \altaffilmark{4} }
\affil{Universit\'e C\^ote d'Azur, Observatoire de la C\^ote d'Azur, CNRS, UMR 7293 Laboratoire Lagrange, 28 Av. Valrose, 06108 Nice Cedex 2, France}
\email{Armando.Domiciano@oca.eu}

\begin{abstract}
We consider a binary system composed of a pulsar and a massive, fast rotating, highly distorted main sequence star of mass $M$, spin angular momentum $\bds S$, dimensionless mass quadrupole moment $J_2$,  equatorial and polar radii $R_\textrm{e},~R_\textrm{p}$, flattening $\nu\doteq (R_\textrm{e}-R_\textrm{p})/R_\textrm{e}$, and ellipticity $\varepsilon\doteq\sqrt{1-R_\textrm{p}^2/R_\textrm{e}^2}$ as a potential scenario to dynamically put to the test certain post-Keplerian  effects of both Newtonian and post-Newtonian nature. We numerically produce time series of the perturbations $\Delta\ton{\delta\tau}$ of the R{\o}mer-like, orbital component of the pulsar's time delay $\delta\tau$  induced over 10 years by the pN gravitoelectric mass monopole $\ton{\textrm{Schwarzschild},~GMc^{-2}}$, quadrupole $\ton{GMR^2_\textrm{e}J_2 c^{-2}}$, gravitomagnetic spin dipole $\ton{\textrm{Lense-Thirring},~GSc^{-2}}$ and octupole $\ton{GSR^2_\textrm{e}\varepsilon^2 c^{-2}}$ accelerations along with the Newtonian quadrupolar $\ton{GMR^2_\textrm{e}J_2}$ one. We do not deal with the various propagation time delays due to the travelling electromagnetic waves.
It turns out that, for a Be-type star with $M = 15~\textrm{M}_\odot,~R_\textrm{e} = 5.96~\textrm{R}_\odot,~\nu = 0.203,~S = 3.41\times 10^{45}~\textrm{J}~\textrm{s},\,J_2 = 1.92\times 10^{-3}$ orbited by a pulsar with an orbital period $\Pb\simeq 40-70~\textrm{d}$,
the classical oblateness-driven effects are at the $\lesssim 4-150~\textrm{s}$ level, while the pN shifts are of the order of
$
\lesssim 1.5-20~\textrm{s}~\ton{GMc^{-2}},~
\lesssim 10-40~\textrm{ms}~\ton{GMR^2_\textrm{e} J_2 c^{-2}},~
\lesssim 0.5-6~\textrm{ms}~\ton{GSc^{-2}},~
\lesssim 5-20~\upmu\textrm{s}~\ton{GSR^2_\textrm{e}\varepsilon^2 c^{-2}}
$, depending on their orbital configuration. The root-mean-square (rms) timing residuals $\upsigma_{\tau}$ of almost all the existing non-recycled, non-millisecond pulsars orbiting massive, fast rotating main sequence stars
%(PSR B1259-63, PSRJ0045-7319, PSR J1638-4725, PSR J1740-3052, PSR %J2032+4127)
are  $\lesssim\textrm{ms}$.
Thus, such kind of binaries have the potential to become interesting laboratories to measure, or, at least, constrain, some Newtonian and post-Newtonian ($GMc^{-2},\,GMJ_2c^{-2}$, and, perhaps, $GSc^{-2}$ as well) key features of the distorted gravitational fields of the fast rotating stars hosted by them.
\end{abstract}

keywords{
gravitation $-$ binaries: general $-$ stars: rotation $-$  pulsars: general $-$ celestial mechanics
}
%keywords{Astrophysical studies of gravity; General relativity; Cosmological constant; Neutron stars \& pulsars; Classical black holes}
%
\section{Introduction}\lb{intro}
In its weak-field and slow-motion approximation, general relativity predicts that, in addition to the time-honored post-Newtonian (pN) gravitoelectric and gravitomagnetic   precessions induced by the mass $M$ (Schwarzschild) and the spin angular momentum $\bds S$ (Lense-Thirring)  of the central body acting as source of the gravitational field, other  pN gravitoelectric and gravitomagnetic orbital effects related to its oblateness  arise as well \citep{Sof89,1988CeMec..42...81S,1990CeMDA..47..205H,1991ercm.book.....B,2014PhRvD..89d4043W,2014CQGra..31x5012P,2015IJMPD..2450067I,2015CeMDA.123....1M,2016JGeod..90.1345S,2018RSOS....580640F,2018CeMDA.130...40S}. So far, they have never been put to the test in any astronomical and astrophysical scenarios, despite some recent preliminary investigations pertaining the planet Jupiter in our solar system \citep{2013CQGra..30s5011I,2019MNRAS.484.4811I}; for some embryonic thoughts about an Earth-spacecraft scenario, see \citet{2013CQGra..30s5011I,2015IJMPD..2450067I}. To the pN level, the oblateness of astronomical bodies modifies also the propagation of the electromagnetic waves in their deformed spacetime. About the perspectives of measuring the resulting deflection due to Jupiter with astrometric techniques, see, e.g.,
\citet{2006CQGra..23.4853C},\,\citet{2007PhRvD..75f2002K},\,\citet{2008PhRvD..77d4029L},\,\citet{2019MNRAS.485.1147A}, and references therein.

An analysis of the analytical expressions of the pN gravitoelectric and gravitomagnetic orbital precessions due to the asphericity of the primary \citep{2015IJMPD..2450067I,2019MNRAS.484.4811I} shows that the key ingredients  needed to enhance their magnitude are a strongly distorted central body, and a highly eccentric and close orbit of the moving particle.

Binaries composed by a pulsar and a main-sequence star \citep{1998MNRAS.298...67W} may offer, in principle, interesting natural laboratories to try to investigate such little known pN effects.
Indeed, they are systems composed of a neutron star regularly emitting electromagnetic radio pulses orbiting an usually more massive main sequence star, which, in most cases, is highly distorted due to its fast rotation, along a generally elliptical orbit.
A very accurate observable quantity in binary pulsars is represented by the measurement of the times of arrival (TOAs) $\tau$ of the pulses emitted by the neutron star which, in case it has a gravitationally bound companion, exhibit a regular variation $\delta\tau$ due to, among other things, the Keplerian motion about the common centre of mass: it is the R{\o}mer-like time delay. The full variation of the pulses' times of arrival is due to several other effects connected, e.g., with the propagation of the electromagnetic waves through the deformed spacetime of the system \citep{1998MNRAS.298...67W}.

The first binary pulsar hosting a main sequence star to be discovered was PSR B1259-63 \citep{1992ApJ...387L..37J,2014MNRAS.437.3255S}; it is characterized by a highly eccentric orbit ($e=0.870$) with an orbital period of $\Pb=1237~\textrm{d}=3.38~\textrm{yr}$. The pulsar's non-degenerate companion is the fast spinning Be star LS 2883, whose equatorial velocity $V_\textrm{e}$ is about $280~\textrm{km~s}^{-1}$ corresponding to $\sim 70\%$ of its break-up velocity \citep{1996MNRAS.280L..31P}. Its mass $M$ and equatorial radius $R_\textrm{e}$ amounts to about 30 Solar masses $(\textrm{M}_\odot)$ and $9.7$ Solar radii $(\textrm{R}_\odot)$ \citep{negueruela+11}.
Later, the eccentric ($e=0.808$) binary PSR J0045-7319 was discovered \citep{1994ApJ...423L..43K}. To date, it is the fastest orbiting system since it is $\Pb=51.17~\textrm{d}$. Its primary is a  main sequence B-star spinning close to its break-up velocity \citep{1995ApJ...452..819L,1996Natur.381..584K}. PSR J1638-4725, having an orbital period of $\Pb=1940~\textrm{d}=5.3~\textrm{yr}$ and $e=0.95$, was found by \citet{2006MNRAS.372..777L}. Its stellar companion should be a rapidly rotating Be star. PSR J1740-3052 \citep{2001MNRAS.325..979S}, with $\Pb=231~\textrm{d}$ and $e=0.578$, hosts most likely a B-type main-sequence star \citep{2010MNRAS.406.1848T,2012MNRAS.425.2378M}. The most recently discovered main-sequence-star binary pulsar is the highly eccentric ($e=0.93$) PSR J2032+4127 \citep{2015MNRAS.451..581L} characterized by $\Pb=8578~\textrm{d}=23.5~\textrm{yr}$. The companion of the neutron star is the massive Be star MT91 213 with $M\simeq 15\,\textrm{M}_\odot$.

For the sake of completeness, we mention also a few other  binary pulsars hosting a non-degenerate star, although they are not relevant for our purposes in view of the nature of their non massive and fast-rotating partners. They are PSR J1903+0327 \textcolor{black}{\citep{2018ApJS..235...37A}}, whose companion is a F5V-GOV $\sim 1~\textrm{M}_\odot$ star moving in $\Pb=95~\textrm{d}$ along a rather eccentric orbit with $e=0.44$, the transitional millisecond pulsar PSR J1023+0038 \citep{2009Sci...324.1411A} orbiting a low-mass ($0.2~\textrm{M}_\odot$) companion star in a circular path with $\Pb=4.75~\textrm{hr}$.
%
%, and PSR J2051-0827 \citep{2001MNRAS.321..576S} whose light companion  $(M=0.05\,\textrm{M}_\odot)$ orbits it in $\Pb = 2.37~\textrm{hr}$ along a tight %circular orbit.
%
The rms timing residuals of the aforementioned binary pulsars are all of the order of $\lesssim \textrm{ms}$, apart from PSR J1903+0327
%
%and PSR J2051-0827
%
which \textcolor{black}{is} at the $\simeq \textcolor{black}{1}\,\upmu\textrm{s}$ level; more specifically, they are $\simeq 0.46~\textrm{ms}$  over $13~\textrm{yr}$ for PSR B1259-63 \citep{2004MNRAS.351..599W}, $7.4~\textrm{ms}$ over $2~\textrm{yr}$ for PSR J0045-7319 \citep{1994ApJ...423L..43K}, $\simeq 5.3~\textrm{ms}$  over $4.35~\textrm{yr}$ for PSR J1638-4725 \citep{2006MNRAS.372..777L}, $\simeq 0.8~\textrm{ms}$  over $2.29~\textrm{yr}$ for PSR J1740-3052 \citep{2001MNRAS.325..979S}, $\simeq 0.5-1~\textrm{ms}$  over about $6~\textrm{yr}$ for PSR J2032+4127 \citep{2015MNRAS.451..581L}, $\simeq 1~\textrm{\upmu s}$  over about $3~\textrm{yr}$ for PSR J1903+0327 \textcolor{black}{\citep{2018ApJS..235...37A}},  $0.1~\textrm{ms}$  over about $4~\textrm{yr}$ for PSR J1023+0038 \citep{2013arXiv1311.5161A}.
%
% and $12.2\,\upmu\textrm{s}$ after $13\,\textrm{yr}$ for PSR J2051-0827 \citep{2011MNRAS.414.3134L}.
%
Table~\ref{tavola1} summarizes the key data for the binary pulsars hosting massive, fast rotating main sequence stars.
\begin{sidewaystable}[ht]
\begin{center}
\small{
\begin{tabular}{|l|l|l|l|l|l|l|l|l|l|l|}
  \hline
Pulsar  &  Companion  & Distance (kpc) &  $M~(\textrm{M}_\odot)$ & $R_\textrm{e}~(\textrm{R}_\odot)$ & $R_\textrm{p}~(\textrm{R}_\odot)$ & $P_\textrm{b}$ (d) &  $e$ & $P$ (ms)  & $\upsigma_\tau$ (ms) &  $\Delta T$ (yr) \\
\hline
PSR J0045-7319 & B1V star & In SMC & $8.8\pm 1.8$ & $6.4\pm 0.7$ & $-$ &  $51.17$ & $0.808$ & $930$ & $7.4$ & $2$ \\
PSR J1740-3052 & Main sequence star & $-$ & $> 11$ & $-$ & $-$ & $231$ & $0.578$ & $570$ & $0.8$ & $2.29$ \\
PSR B1259-63 & B2e star LS 2883 & $2.75$ & $\simeq 30$ & $\simeq 9.7$ & $\simeq 8.1$ &  $1237$ & $0.870$ & $48$ & $0.46$ & $13$ \\
PSR J1638-4725 & Be star & $-$ &  $>4$ & $-$ & $-$ & $1940$ & $0.95$ & $764$ & $5.3$ & $4.35$ \\
PSR J2032+4127 & Be star MT91 213 & $1.7$ & $\simeq 15$ & $-$  & $-$  & $8578$ & $0.93$ & $143$ & $0.5-1$ & $6$ \\
\hline
\end{tabular}
}
\end{center}
\caption{Binary pulsars with a massive, fast rotating main sequence star companion discovered so far. For each of them, we list the companion, the distance (when available), the mass $M$, the equatorial and polar radii $R_\textrm{e},\,R_\textrm{p}$ (when available), the orbital period $P_\textrm{b}$, the orbital eccentricity $e$, the spin period $P$, the rms timing residuals $\upsigma_\tau$, the data analysis time span $\Delta T$. For the values of the listed parameters, see the references cited in the text and the online pulsar catalog at http://www.atnf.csiro.au/research/pulsar/psrcat/. For PSR B1256-63 we give the equatorial $R_\textrm{e}$ and polar $R_\textrm{p}$ radii as estimated by \citet{negueruela+11}.}\label{tavola1}
\end{sidewaystable}

About the achievable accuracy level in timing residuals, in the case of the pulsars orbiting a main sequence star, their timing seems doomed to stay at the $\simeq\textrm{ms}$ level. It is so because, for evolutionary reasons, they are not fully recycled \citep{2010NewAR..54...93S}. Thus, their spinning periods $P$ are not at the millisecond level, and their TOAs are not measured with a precision of the order of $\simeq \upmu\textrm{s}$\textcolor{black}{, and their timing is often contaminated by timing noise \citep{Hobbsetal010}}. The timing of the non-recycled pulsars is almost always less accurate than for the fully recycled pulsars. Indeed, PSR J1903+0327 rotates with a period $P=2.5\,\textrm{ms}$, and its rms timing residuals are as little as $\simeq 1\,\upmu\textrm{s}$. The spin period of PSR J1023+0038 is $P=1.7\,\textrm{ms}$, and its rms timing residuals are  $0.1\,\textrm{ms}$.
%
%Also PSR J2051-0827 is a millisecond pulsar $(P=4.5\,\textrm{ms})$, and its decadal rms timing residuals is at the $\simeq 10\,\upmu\textrm{s}$ level.
%
Incidentally, we mention the fact that, according to Table\,2  of
\citet{2018ApJS..235...37A}, the rms timing accuracy of some fully recycled pulsars with $P=1.5-10\,\textrm{ms}$, isolated or with a white dwarf as companion, is of the order of $0.1-0.2\,\upmu\textrm{s}$. It is expected that future instrumental improvements may push the rms timing accuracy of some of them to the\footnote{A. Possenti, personal communication to L.I., April 2019.} $\simeq 10\,\textrm{ns}$ level over time spans some yr long. For binary pulsars hosting another neutron star, the rms timing accuracy is of the order of $1-100\,\upmu\textrm{s}$.

Here, we will preliminarily investigate the size and the temporal patterns of the perturbations $\Delta\ton{\delta\tau}$ induced on the R{\o}mer-like orbital part of the pulsar's time delay $\delta\tau$ by both the standard (Schwarzschild and Lense-Thirring) and the oblateness-driven pN accelerations felt by a fictitious neutron star orbiting a highly distorted, fast rotating B-type main sequence star in view of a possible detection in new binaries that may eventually be discovered in the future.
However, caution is in order before inferring too optimistic conclusions from a  straightforward comparison of our simulated time series with the rms timing residuals listed in Table\,\ref{tavola1}. Even if the size of some pK signatures were to be larger than the $\simeq\textrm{ms}$ level, it does not necessarily mean that such  effects will be measurable in the actual processing of the real observations. Indeed, careful, dedicated simulated data reductions and covariance analyses should be performed by explicitly modeling the signals of interest, estimating its characteristic parameters and inspecting the resulting correlations with the other parameters usually estimated. It should be kept in mind that, in principle, an unmodeled effect may be removed from the post-fit residuals, at least to a certain extent, being \virg{absorbed} in the estimated values of the other parameters determined in the data reduction.
Thus, our investigation should be deemed just as a sensitivity analysis able to preliminarily explore the potential of the scenarios considered.
We will assume the validity of general relativity throughout the paper, which is organized as follows.

In Section~\ref{J2S}, we discuss the magnitude of the angular momentum $S$ and the dimensionless quadrupole mass  moment $J_2$ of typical fast rotating massive B-type stars. Section~\ref{GR} is devoted to the numerical calculation of the  perturbations $\Delta\ton{\delta\tau}$ induced on the pulsar's R{\o}mer time delay by some post-Keplerian (pK) classical and pN  accelerations. We do not calculate the propagation delays accounting, e.g., for the effects on the pulsar's travelling electromagnetic waves through the deformed spacetime of the B-type star. Section~\ref{conclu} summarizes our findings and contains our conclusions. Once again, we stress the preliminary nature of our sensitivity investigation; we do not perform a full covariance analysis implying, e.g., the simulation of the pulsar's TOAs and their reduction along with parameter estimation.
\section{Quadrupole mass moment, flattening, and angular momentum of fast rotating main sequence B-type massive stars}\lb{J2S}

The binary pulsars on which we shall focus presumably own a main sequence B-type
massive star. Such stars are mostly fast rotators \cite[][]{levato+13} and
are hence distorted by the centrifugal acceleration. Such a
distortion takes the mass distribution away from spherical symmetry
and endows these stars with a quadrupolar and higher gravitational
moments. In the case of Be stars (namely B stars with emission lines),
rotation is believed to be almost critical, namely
the rotational velocity is taken close ($> 70\%$) to the Keplerian velocity at equator \cite[e.g.][]{Rivinius2013_v21p69}. When
critical rotation is reached, the surface distortion is maximum and
the flattening $\nu\doteq (R_\textrm{e}-R_\textrm{p})/R_\textrm{e}$\textcolor{black}{, expressed in terms of the equatorial and polar radii $R_\mathrm{e},\,R_\mathrm{p}$, respectively,} is close to one third. For such
stars the computation of the gravitational quadrupolar moment cannot
be done perturbatively as it is the case for the Sun, which is a slow
rotator \cite[e.g.][]{rozelot_etal01}. Modelling these stars requires
two-dimensional models. Fortunately, self-consistent 2D-models have recently been
achieved with the ESTER code \cite[][]{ELR13,RELP16}. Compared to previous
2D-models, ESTER models include self-consistently the baroclinicity of
the stellar envelopes and can thus predict the associated differential
rotation. They therefore provide unambiguously the total angular momentum
of the star given, for instance, its equatorial velocity.

With the ESTER code, we computed the parameters of three stellar models of 10, 15, and 30\,$\textrm{M}_\odot$ as they can represent the companions of PSR J0045-7319, PSR J2032+4127 and PSR B1256-63 respectively. Since the evolutionary status of the stars is unknown but presumably on or close to the main sequence, we computed their steady state at ZAMS (Zero-Age Main Sequence) and at half-main sequence to get an idea of the effects of evolution. We use standard galactic metallicity $Z=0.02$ with the solar mixture (which may be approximate for PSR J0045-7319 which is in the SMC, known to be less metallic than the Galaxy). Results of ZAMS and evolved models are displayed in Tables~\ref{Ester_zams}\,to\,\ref{Ester_ev} respectively. There we give the total spin angular momentum $S$ and the dimensionless quadrupole mass moment $J_2$ along with other bulk parameters of the models. We recall that multipole gravitational moments $J_\ell$ are defined by the multipole expansion of the gravitational potential of a mass $M$, namely
\begin{equation}
U\ton{\bds r} = -\frac{\mu}{r}\left[1 - \sum_\ell J_\ell\left(\frac{R_\textrm{e}}{r}\right)^\ell \mathcal{P}_\ell\ton{\xi}\right],\lb{potenz}
\end{equation}
where $\mu\doteq GM$ is the star's gravitational parameter, $G$ is the Newtonian constant of gravitation, $\xi\doteq \kap\bds\cdot\bds{\hat{r}}$ is the cosine of the angle $\theta$ between the directions of the body's spin axis and of an external point at $\bds r$,  $\mathcal{P}_\ell\ton{\xi}$ is the Legendre polynomial of degree $\ell$.
We consider the mass distribution of the stellar models to be symmetric with respect to equator thus making the $J_\ell$ of odd order all vanish. The remaining $J_{2p}$ can be computed with the integral expression
\begin{equation}
J_{2p} = -\frac{1}{MR^{2p}_\textrm{e}}\int_{\mathcal{V}} r^{2p}\,\mathcal{P}_{2p}\ton{\xi}\,\rho\ton{\bds r}\,d^3\bds r
\end{equation}
where the integration is over the volume $\mathcal{V}$ of the star. The same expression is given
in, for instance, \cite{helled+11}. Here we are especially interested in $J_2$ and $S$, namely in
\begin{equation}
J_2 = -\frac{1}{MR^2_\textrm{e}}\int_{\mathcal{V}} r^2\,\mathcal{P}_2\ton{\xi}\,\rho\ton{\bds r}\,d^3\bds r\quad {\rm and}\quad S =  \int_{\mathcal{V}}r^2(1-\xi^2)\,\Xi\ton{\bds r}\rho\ton{\bds r}\,d^3\bds r
\end{equation}
which are directly computed from the ESTER models; $\Xi\ton{\bds r}$ is the local angular speed.

We choose an angular rotation rate of 70\% of the actual critical (Keplerian) angular velocity of the star. Such a rotation rate is typical of the nearby fast rotating stars that have been measured by interferometry. Their flattening is typically $\sim0.2$  \cite[e.g.][]{domiciano_etal14}, as our models.
\begin{table}[ht]
\caption{ESTER models for Zero-Age Main Sequence (ZAMS) stars with an equatorial angular velocity at
70\% of the critical angular velocity. $R_\textrm{e}$ and $V_\textrm{e}$ are the equatorial radius and
velocity, respectively, $S$ is the total spin angular momentum, $\nu$ is the
flattening, and $J_2$ is the dimensionless mass quadrupole  moment. The metallicity is Z=0.02 and the hydrogen mass fraction is $X=0.7$. In our simulations, we use the parameters of the star with $15\,\textrm{M}_\odot$ listed here.}
\label{Ester_zams}
\begin{center}
\begin{tabular}{|l|l|l|l|l|l|}
\hline
$M~(\textrm{M}_\odot)$ & $R_\textrm{e}~(\textrm{R}_\odot)$ & $V_\textrm{e}~(\textrm{km}~\textrm{s}^{-1})$ & $S~(\times 10^{44}\textrm{J}~\textrm{s})$ & $\nu$ & $J_2~(\times 10^{-3})$\\
\hline
  10    &    4.74        &  444                &  15.3               & 0.201      & 1.63 \\
  15    &    5.96        &  485                &  34.1               & 0.203      & 1.92 \\
  30    &    8.89        &  562                &  125.               & 0.210      & 2.17 \\
\hline
\end{tabular}
\end{center}
\end{table}
\begin{table}[ht]
\caption{Same as in table~\ref{Ester_zams} but for ESTER models of
stars at mid-main-sequence, namely when the hydrogen mass fraction in the convective core is half of the initial one.}
\label{Ester_ev}
\begin{center}
\begin{tabular}{|l|l|l|l|l|l|}
\hline
$M~(\textrm{M}_\odot)$ & $R_\textrm{e}~(\textrm{R}_\odot)$ & $V_\textrm{e}~(\textrm{km}~\textrm{s}^{-1})$ & $S~(\times 10^{44}~\textrm{J}~\textrm{s})$ & $\nu$ & $J_2~(\times 10^{-3})$\\
\hline
  10  &    6.93                &  367                &  13.3               & 0.203      & 0.816\\
  15  &    9.07                &  393                &  28.0               & 0.207      & 0.788\\
  30  &    14.8                &  435                &  870.               & 0.227      & 0.495\\
\hline
\end{tabular}
\end{center}
\end{table}

From Tables~\ref{Ester_zams}~to~\ref{Ester_ev}, we clearly see that as evolution proceeds, namely as the hydrogen content of the core decreases, $J_2$ decreases as expected from the resulting contraction of the convective core. From the work of \cite{james64} we can compute $J_2$ for a polytrope of index $n=3$  with a similar flattening as the ESTER models. We find that $J_2\simeq 2.1\times10^{-3}$ which is quite similar to the ZAMS models. For evolved models one should use polytropes with a higher polytropic index, as they are more centrally condensed, and typically $n=3.43$ matches the ESTER evolved models as far as $J_2$ is concerned. This result may be useful for simulating the orbital evolution of binary pulsars since polytropic models are much easier to compute.
\section{The perturbations of the R{\o}mer-type pulsar's time delay due to some pK Newtonian and pN accelerations}\lb{GR}
Here, we will assume a coordinate system centered in the binary's barycenter whose reference $z$-axis is directed along the line of sight from the binary to the observer, while the reference $\grf{x,~y}$ plane spans the plane of the sky.
\subsection{The pulsar as a structureless, pointlike particle}\lb{punto}
We will, first, consider the following pK accelerations experienced by a test particle moving with velocity $\textcolor{black}{\bds{v}}$ in the external field of an oblate body of mass $M$, equatorial and polar  radii $R_\textrm{e},~R_\textrm{p}$, ellipticity $\varepsilon \doteq \sqrt{1- R_\textrm{p}^2/R_\textrm{e}^2}$, angular momentum $\bds S$ and dimensionless quadrupole mass moment $J_2$. In Section\,\ref{esteso}, we will discuss the limits of validity of the point-particle approximation.

To the Newtonian level, the external potential of the distorted star at the position $\bds r$ is, from \rfr{potenz},
\eqi
U\ton{\bds r} = U_0 + \Delta U_2 = -\frac{\mu}{r}\qua{1-\ton{\frac{R_\textrm{e}}{r}}^2 J_2 \mathcal{P}_2\ton{\xi}},
\eqf
where  $\mathcal{P}_2\ton{\xi} = \ton{3\xi^2 - 1}/2$ is the Legendre polynomial of degree 2.
The classical acceleration due to $J_2$ is
\eqi
{\bds A}^{\textrm{N}J_2} = -\bds\nabla \Delta U_{J_2}= \frac{3\mu J_2 R_\textrm{e}^2}{2r^4}\qua{\ton{5\xi^2 - 1}\bds{\hat{r}} - 2\xi\kap}.\lb{NJ2}
\eqf
%where
%\eqi
%\Delta U_{J_2} = \rp{\mu J_2R_\textrm{e}^2}{r^3}P_2\xi.\lb{U2}
%\eqf
%
%It may be interesting to note that, to our knowledge, PSR J2051-0827 is the first and only binary pulsar for which evidence of the quadrupole of the companion %was found \citep{2011MNRAS.414.3134L}, although with a difficult analysis due to the correlations among the several parameters estimated in the data %reduction.

The 1pN gravitoelectric Schwarzschild-like acceleration affecting the motion of a test particle in the static field of a nonrotating, spherically symmetric body is \citep{2010ITN....36....1P}
\eqi
{\bds A}^{\textrm{1pN}M} = \frac{\mu}{c^2 r^2}\qua{\ton{\frac{4\mu}{r}-{\mathrm{v}}^2}\bds{\hat{r}} + 4{\mathrm{v}}_r\textcolor{black}{\bds{v}}},\lb{1pNM}
\eqf
where $c$ is the speed of light in vacuum, and ${\mathrm{v}}_r \doteq \textcolor{black}{\bds{v}}\bds\cdot\bds{\hat{r}}$ is the radial velocity of the test particle. \Rfr{1pNM} is responsible for the formerly anomalous perihelion precession of Mercury whose explanation was the first empirical confirmation of general relativity \citep{1915SPAW...47..831E}.

The 1pN gravitomagnetic Lense-Thirring acceleration in the stationary field due to the rotating primary is \citep{2010ITN....36....1P}
\eqi
{\bds A}^{\textrm{1pN}S} \lb{1pNS}  = \frac{2GS}{c^2 r^3}\qua{3\xi\bds{\hat{r}}\bds\times\textcolor{black}{\bds{v}} + \textcolor{black}{\bds{v}}\bds\times\kap}.
\eqf
The gravitomagnetic field of the Earth was unambiguously measured for the first time by the Gravity Probe B (GP-B) mission \citep{2011PhRvL.106v1101E}. Tests of the Lense-Thirring orbital precessions with some terrestrial geodetic satellites are ongoing; see, e.g. \citet{2013CEJPh..11..531R}, and \citet{2015CQGra..32o5012L} and references therein for comprehensive reviews.

The 1pN gravitoelectric acceleration felt by a test particle in the field of an oblate body  is \citep{1988CeMec..42...81S, Sof89, 1991ercm.book.....B, 2014PhRvD..89d4043W,2015IJMPD..2450067I}
\eqi
{\bds A}^{\textrm{1pN}M J_2} \lb{1pNMJ2} =  \frac{\mu J_2 R_\textrm{e}^2}{c^2 r^4}\grf{\frac{3}{2}\qua{\ton{5\xi^2 - 1}\bds{\hat{r}} - 2\xi\kap  }\ton{{\mathrm{v}}^2 -\frac{4\mu}{r} }- 6\qua{\ton{5\xi^2 - 1}{\mathrm{v}}_r  - 2\xi{\mathrm{v}}_S   }\textcolor{black}{\bds{v}}
- \frac{2\mu}{r}\ton{3\xi^2 - 1}\bds{\hat{r}}},
\eqf
where ${\mathrm{v}}_S \doteq \textcolor{black}{\bds{v}}\bds\cdot\kap$ is the component of the particle's velocity along the direction of the primary's spin.
Note that the parameter $J_2$ in \rfr{1pNMJ2} is the same entering \rfr{NJ2}, as per Equations 1 to 2 of \citet{1988CeMec..42...81S}.

The 1pN gravitomagnetic acceleration imparted to a test particle by the spin octupole moment\textcolor{black}{\footnote{\textcolor{black}{It will be shown that its effects are small enough to justify order-of-magnitude calculations, without need of detailed stellar models.}}} of a uniformly rotating homogenous oblate spheroid  \citep{2014CQGra..31x5012P,2015CeMDA.123....1M,2019MNRAS.484.4811I} can be cast into the compact form
\eqi
\bds A^{\textrm{1pN}SJ_2} = \frac{3GSR_\textrm{e}^2\varepsilon^2 }{7c^2r^5}\textcolor{black}{\bds{v}}\bds\times\grf{5\xi\qua{7\xi^2 - 3}\bds{\hat{r}} + 3\qua{1 - 5\xi^2}\kap}.\lb{1pNSJ2}
\eqf

\begin{figure}[ht]
\begin{center}
\centerline{
\vbox{
\begin{tabular}{cc}
\epsfysize= 4.4 cm\epsfbox{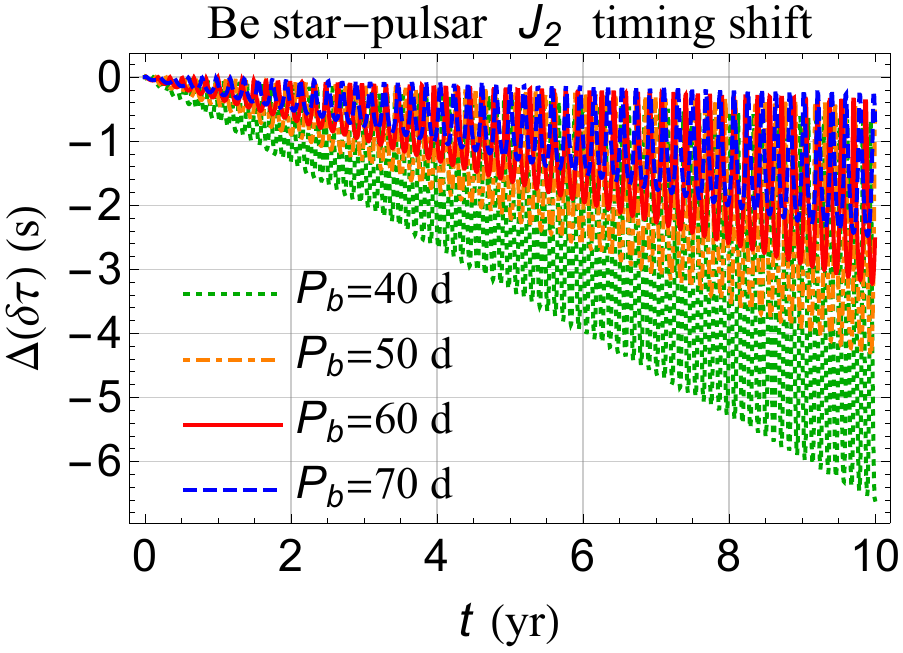}&\epsfysize= 4.4 cm\epsfbox{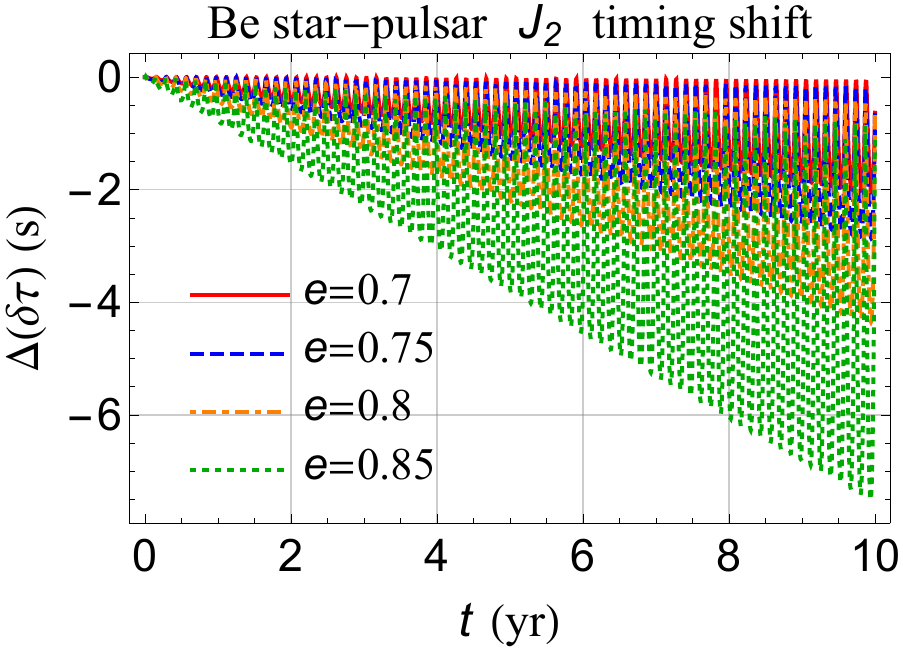}\\
\epsfysize= 4.4 cm\epsfbox{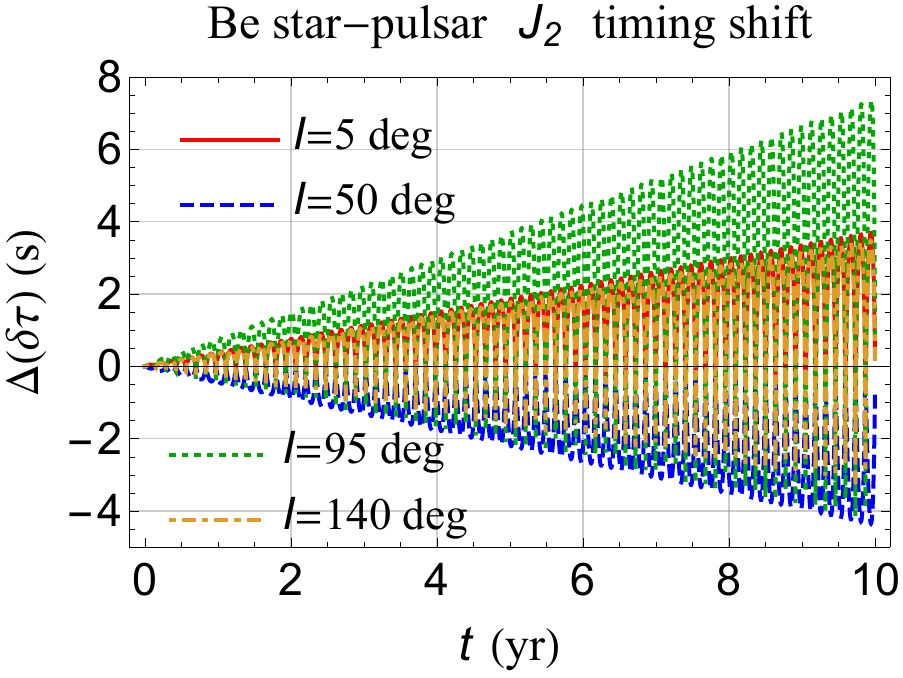}&\epsfysize= 4.4 cm\epsfbox{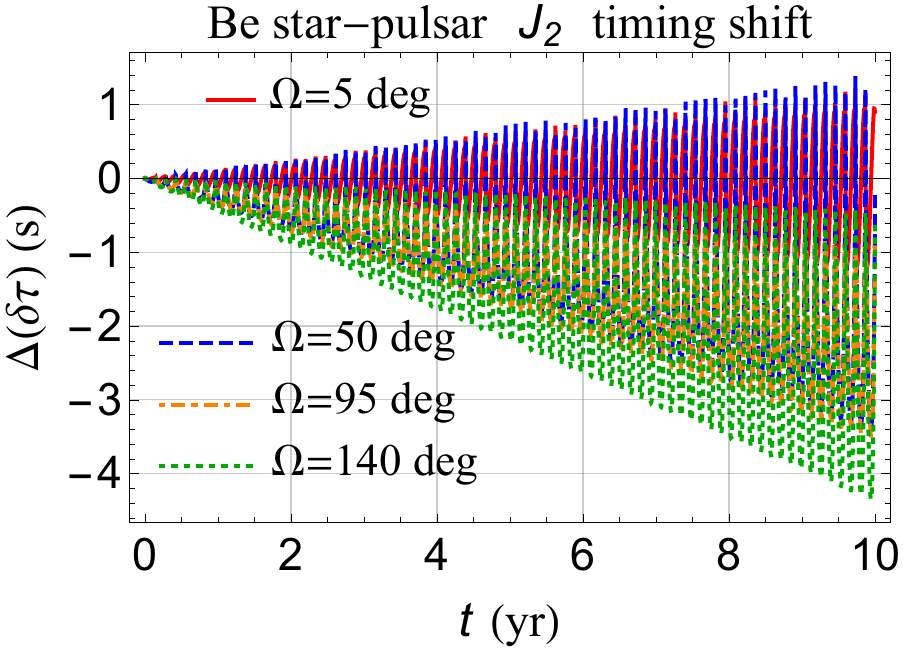}\\
\epsfysize= 4.4 cm\epsfbox{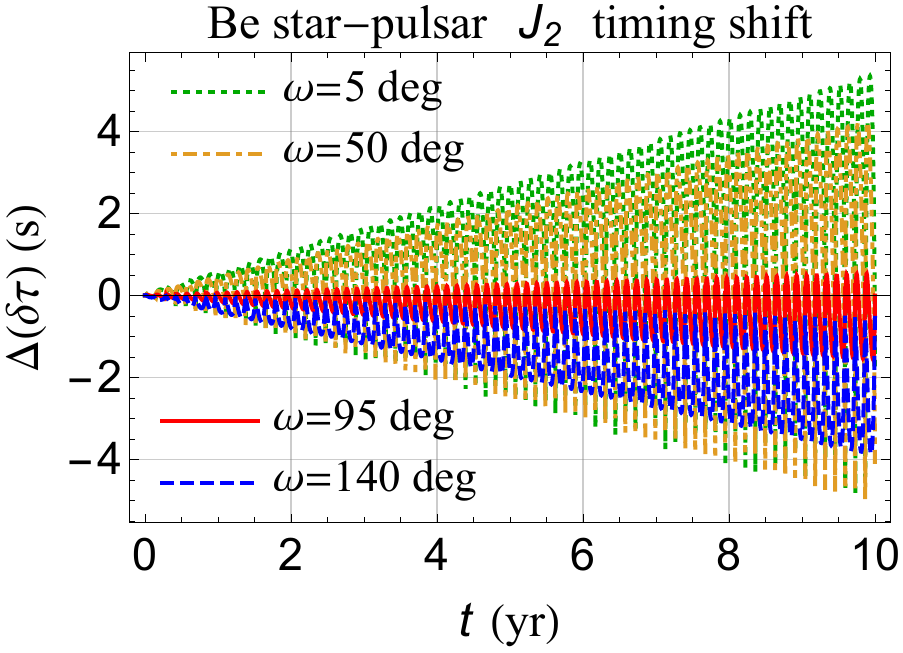}&\epsfysize= 4.4 cm\epsfbox{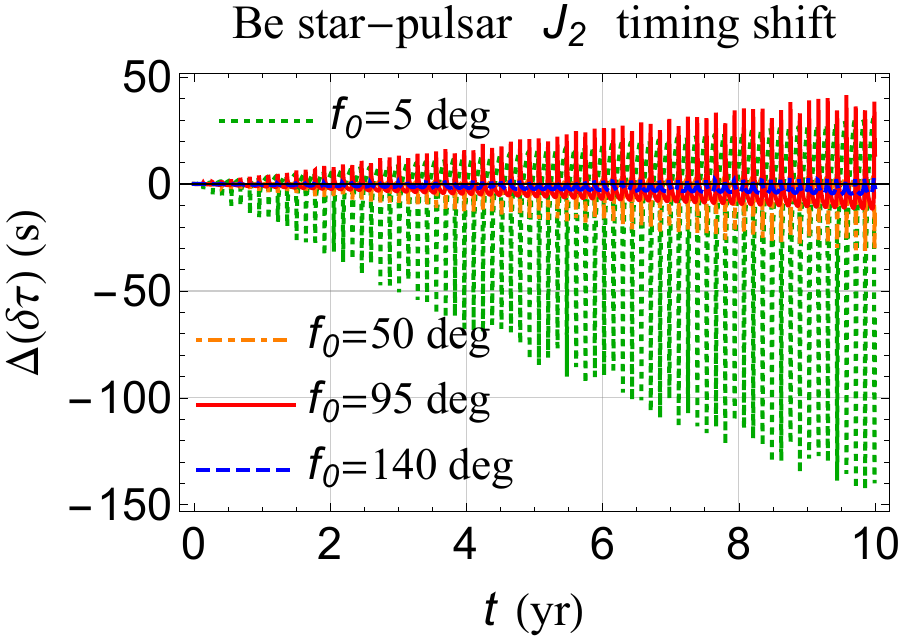}\\
\epsfysize= 4.4 cm\epsfbox{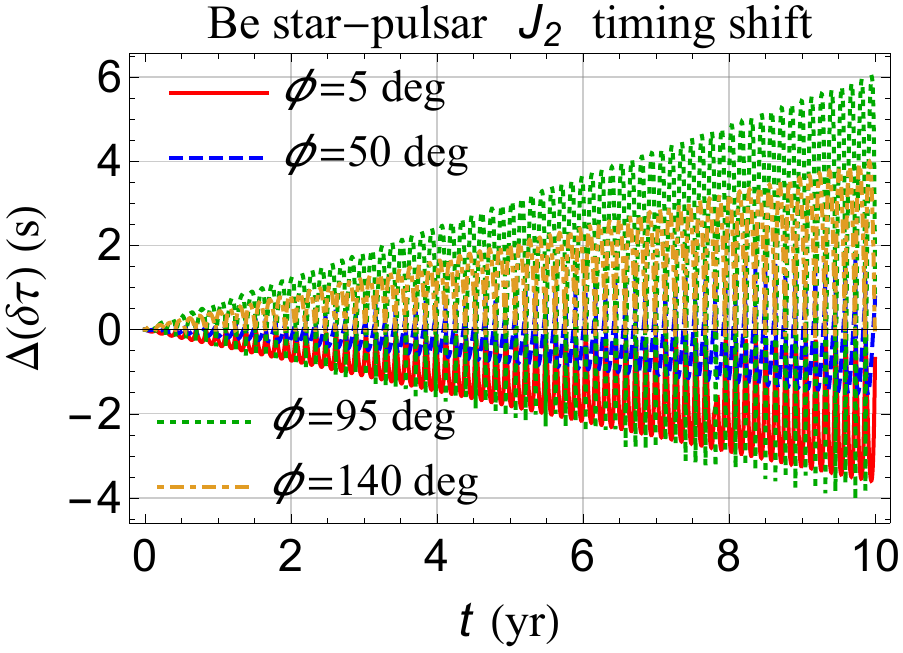}&\epsfysize= 4.4 cm\epsfbox{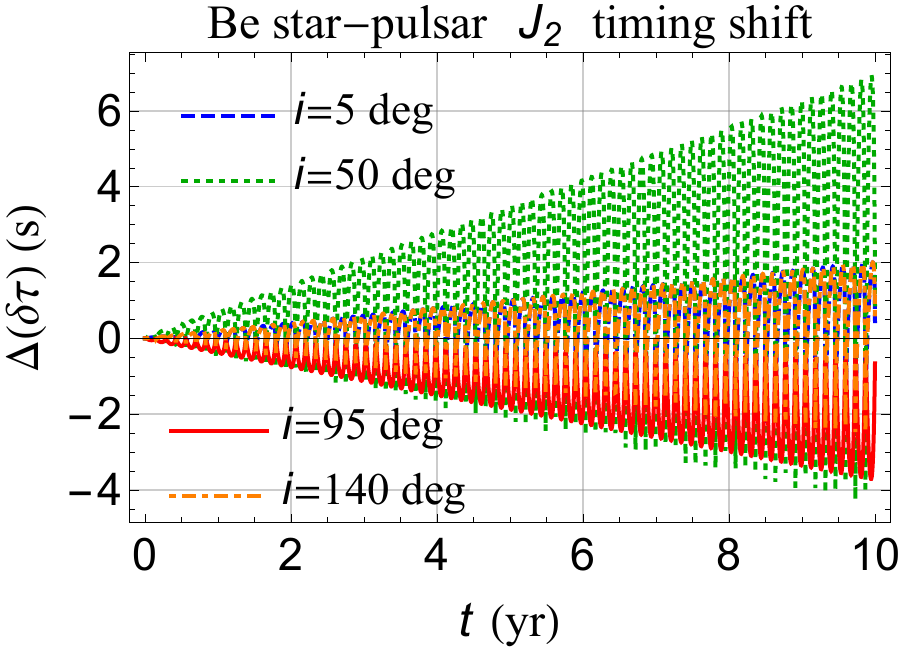}\\
\end{tabular}
}
}
\caption{Numerically integrated time series of the timing shift $\Delta\ton{\delta\uptau}$ due to \rfr{NJ2}, in $\textrm{s}$, for variations of the parameter space of a fictitious binary pulsar, characterized by
$\Pb = 50~\textrm{d},~e=0.8,~I=50~\textrm{deg},~\Omega=140~\textrm{deg},~\omega=149~\textrm{deg},~f_0=228~\textrm{deg}$, orbiting a Be-type star with $M=15~\textrm{M}_\odot,~R_\textrm{e} = 5.96~\textrm{R}_\odot,~J_2 = 1.92\times 10^{-3},~i=60~\textrm{deg},~\phi=217~\textrm{deg}$.}\label{fig_NJ2}
\end{center}
\end{figure}

\begin{figure}[ht]
\begin{center}
\centerline{
\vbox{
\begin{tabular}{cc}
\epsfysize= 4.4 cm\epsfbox{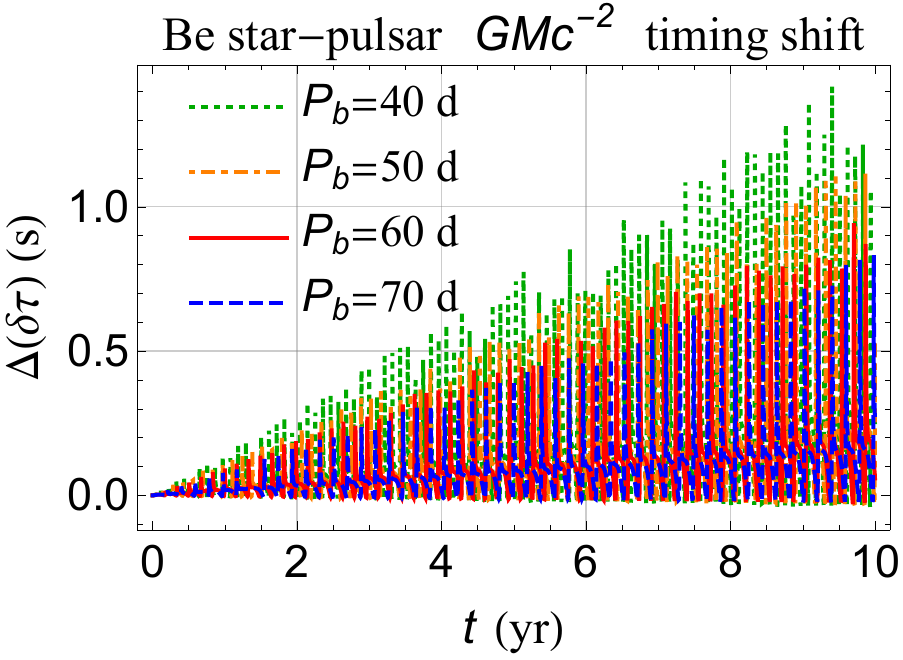}&\epsfysize= 4.4 cm\epsfbox{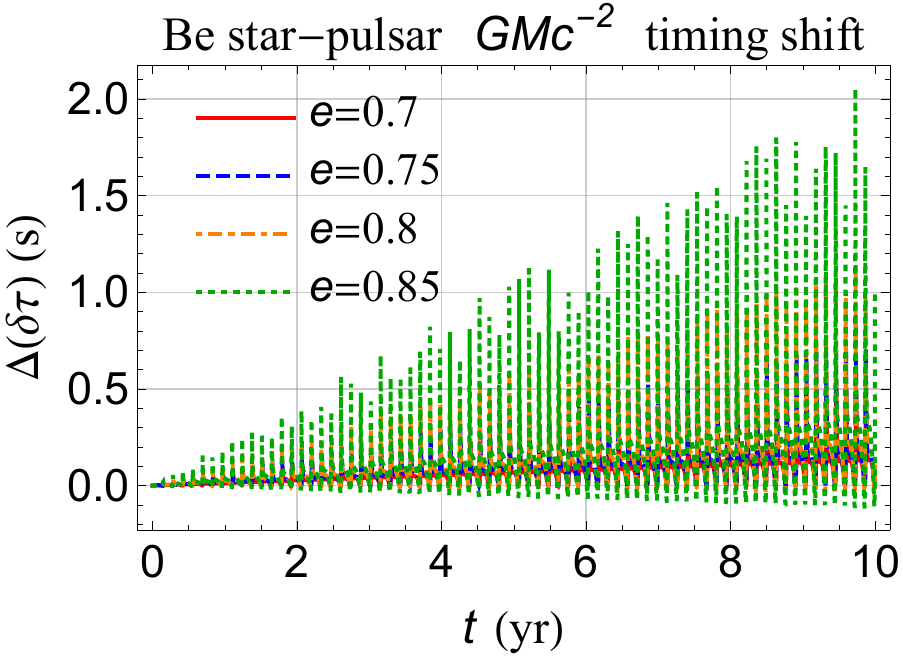}\\
\epsfysize= 4.4 cm\epsfbox{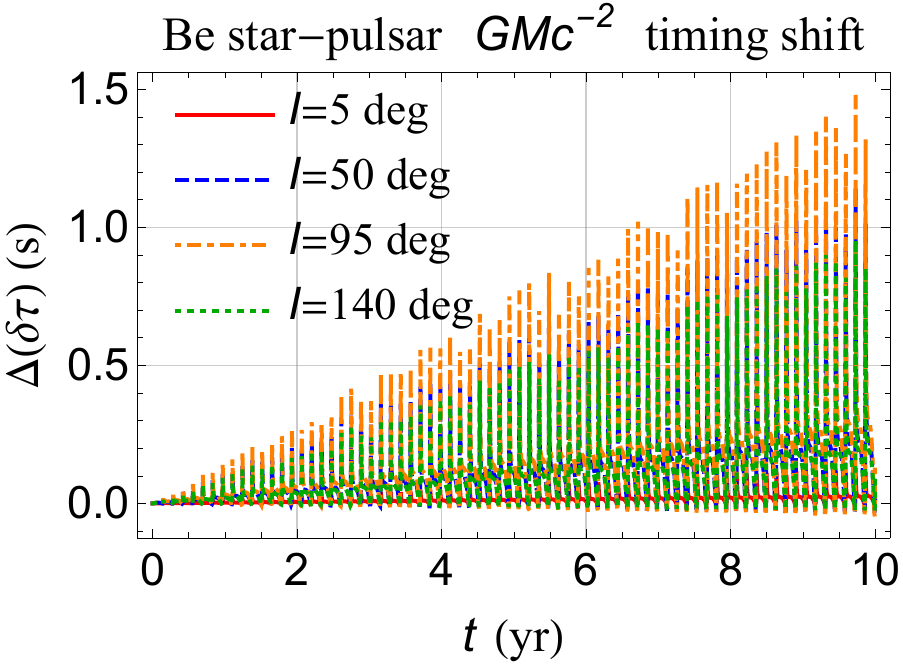}&\epsfysize= 4.4 cm\epsfbox{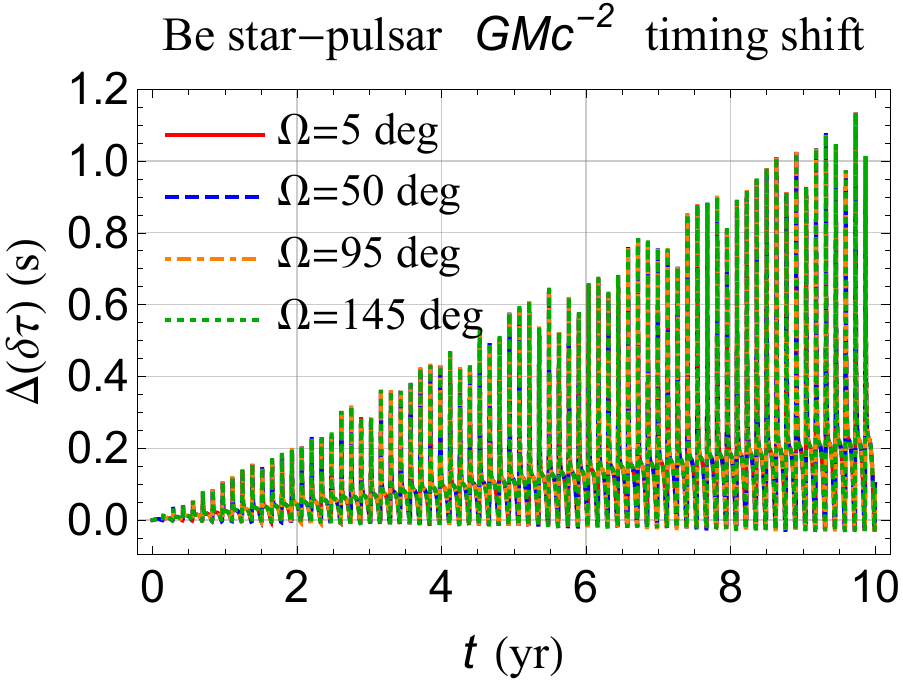}\\
\epsfysize= 4.4 cm\epsfbox{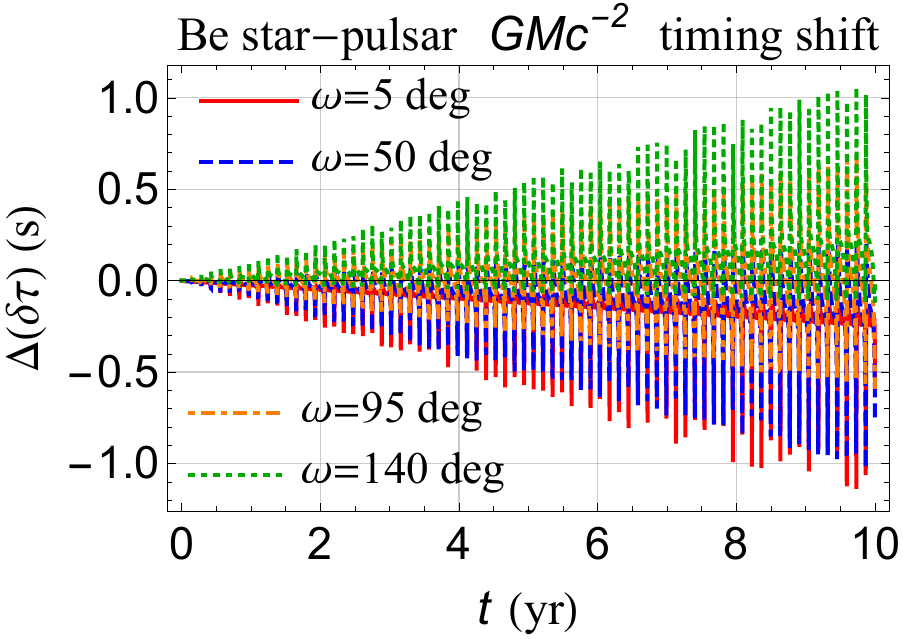}&\epsfysize= 4.4 cm\epsfbox{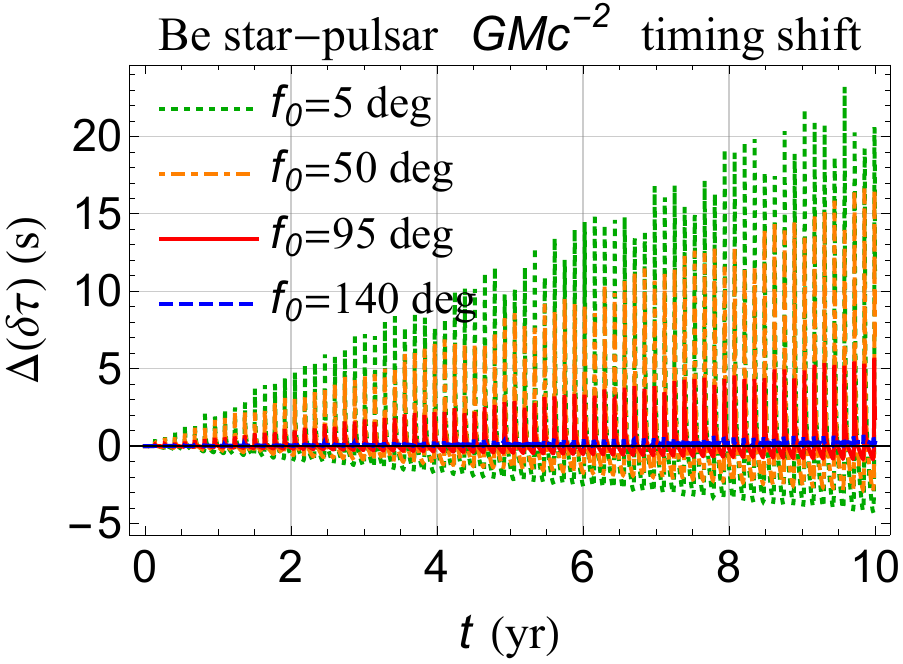}\\
\end{tabular}
}
}
\caption{Numerically integrated time series of the timing shift $\Delta\ton{\delta\uptau}$ due to \rfr{1pNM}, in $\textrm{s}$, for variations of the parameter space of a fictitious binary pulsar, characterized by
$\Pb = 50~\textrm{d},~e=0.8,~I=50~\textrm{deg},~\Omega=140~\textrm{deg},~\omega=149~\textrm{deg},~f_0=228~\textrm{deg}$, orbiting a Be-type star with $M=15~\textrm{M}_\odot$.}\label{fig_1pNM}
\end{center}
\end{figure}

\begin{figure}[ht]
\begin{center}
\centerline{
\vbox{
\begin{tabular}{cc}
\epsfysize= 4.4 cm\epsfbox{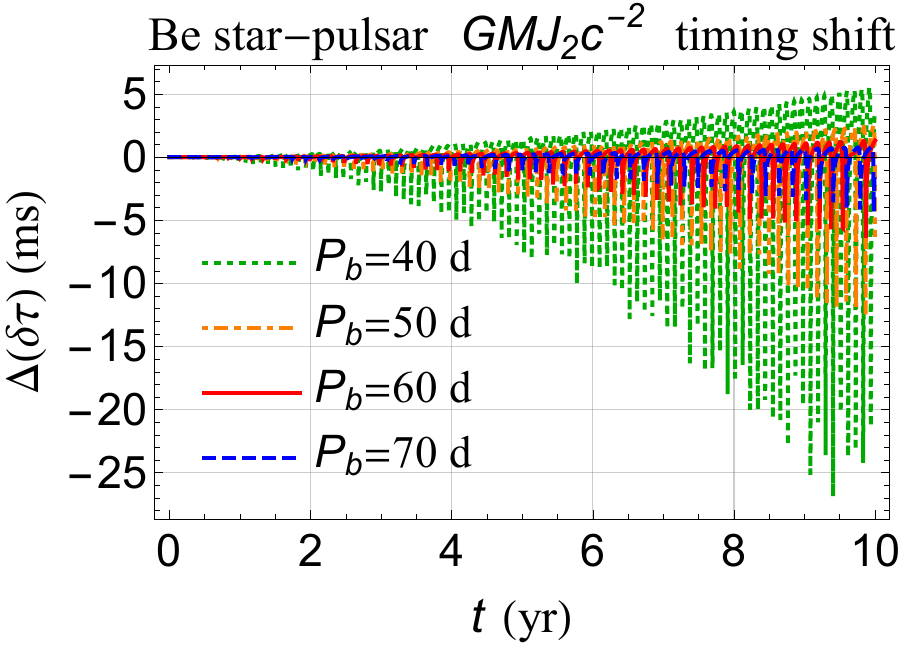}&\epsfysize= 4.4 cm\epsfbox{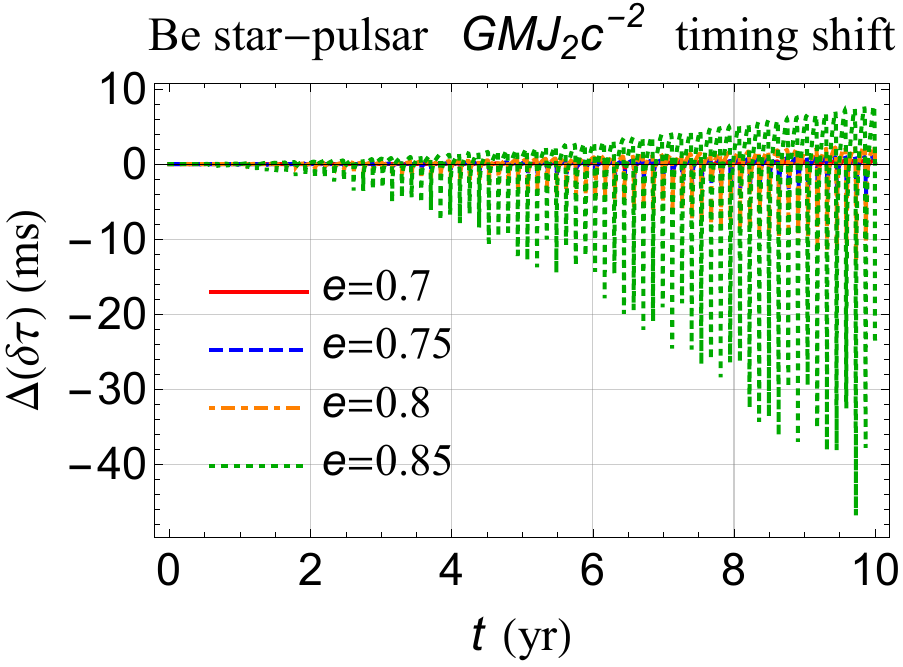}\\
\epsfysize= 4.4 cm\epsfbox{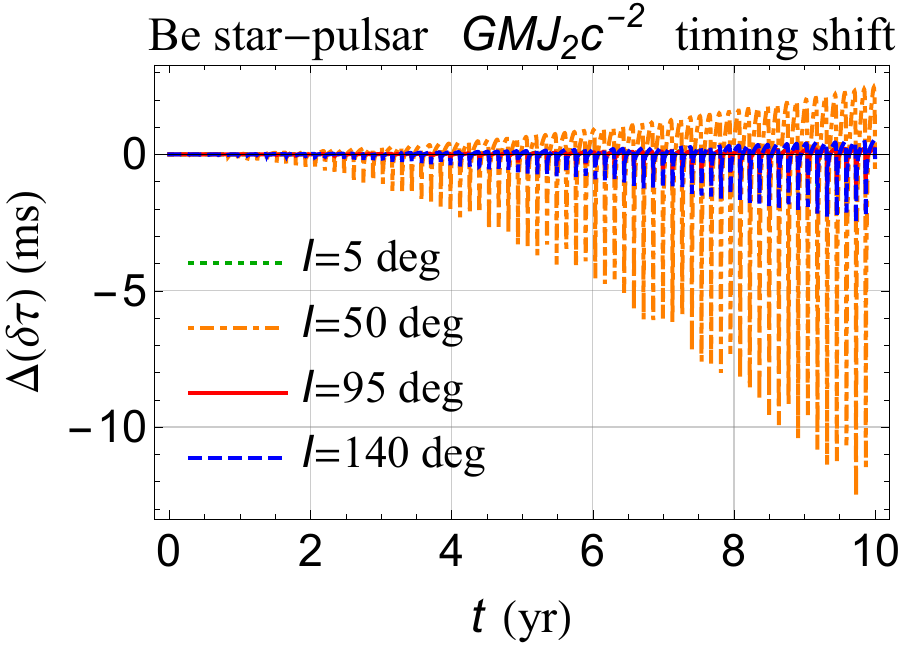}&\epsfysize= 4.4 cm\epsfbox{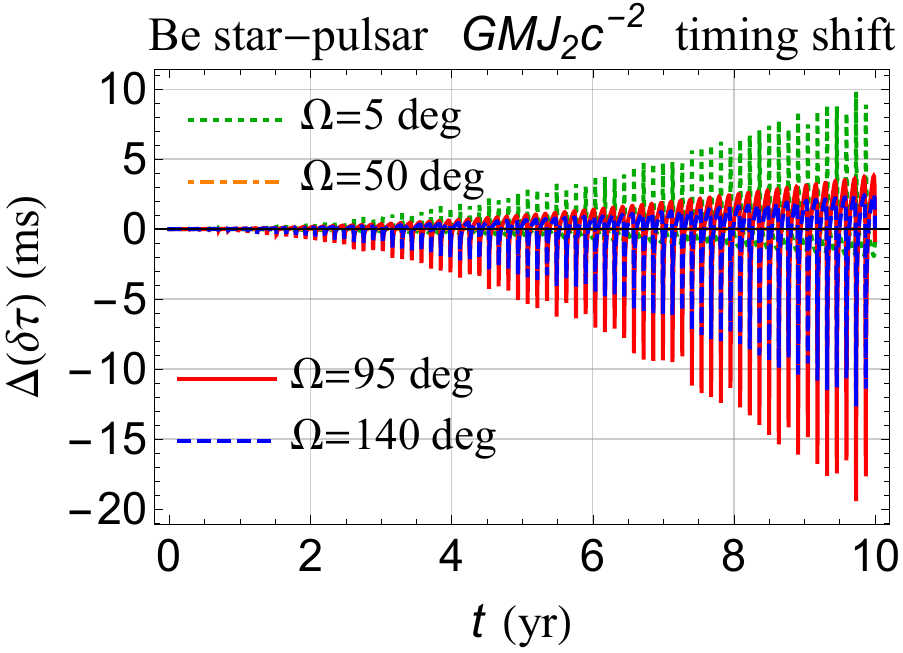}\\
\epsfysize= 4.4 cm\epsfbox{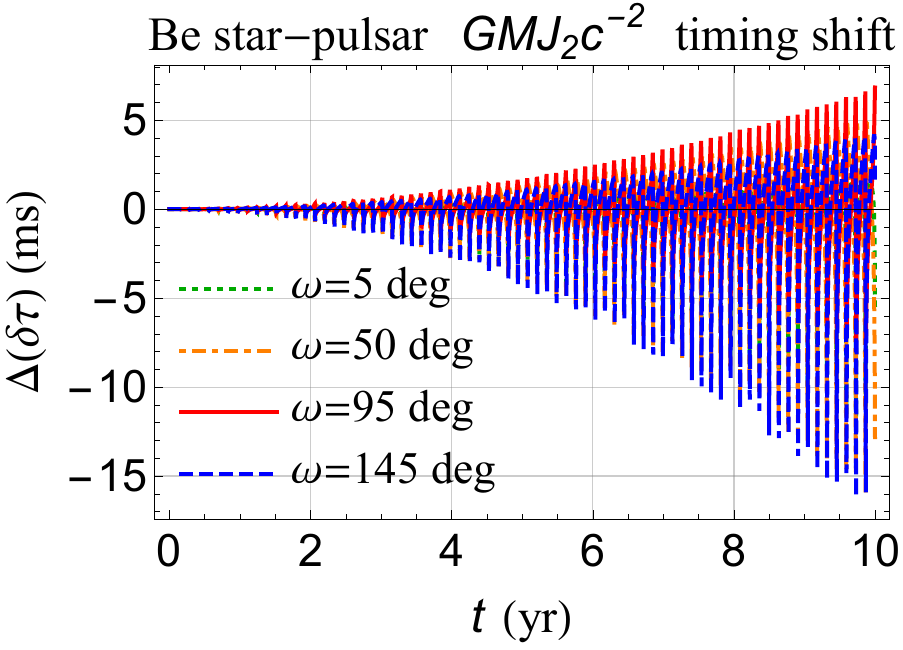}&\epsfysize= 4.4 cm\epsfbox{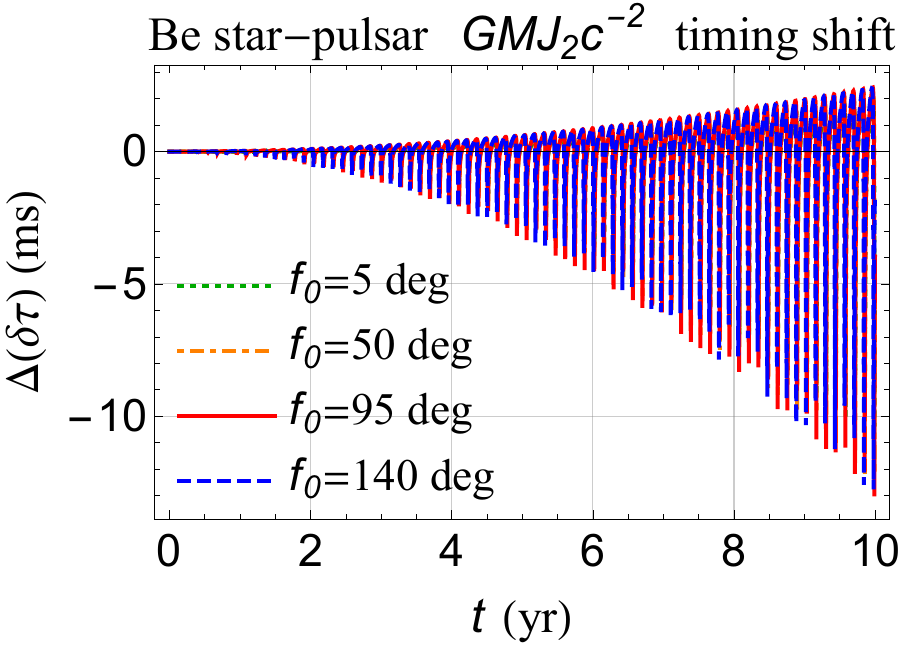}\\
\epsfysize= 4.4 cm\epsfbox{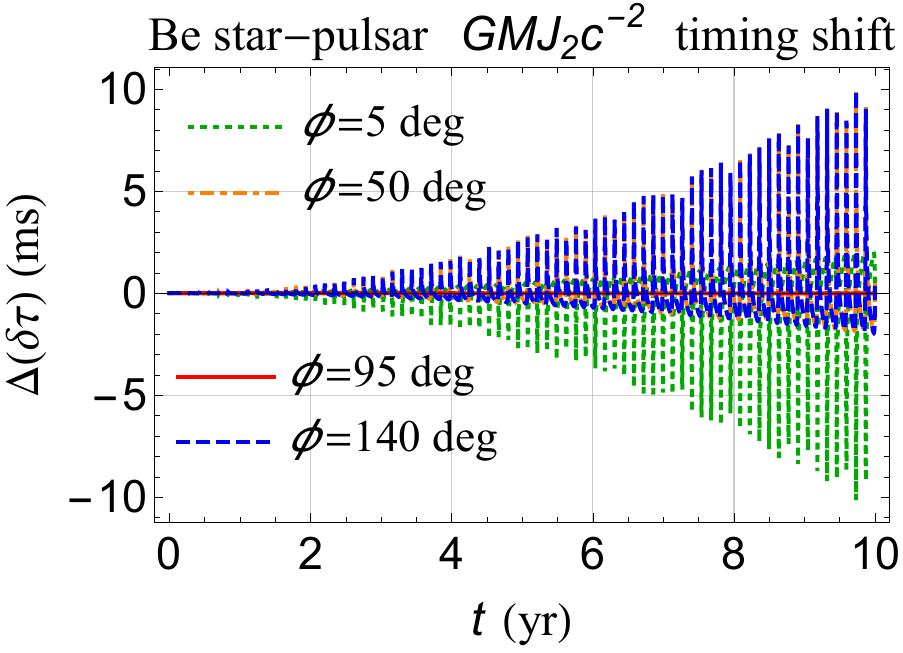}&\epsfysize= 4.4 cm\epsfbox{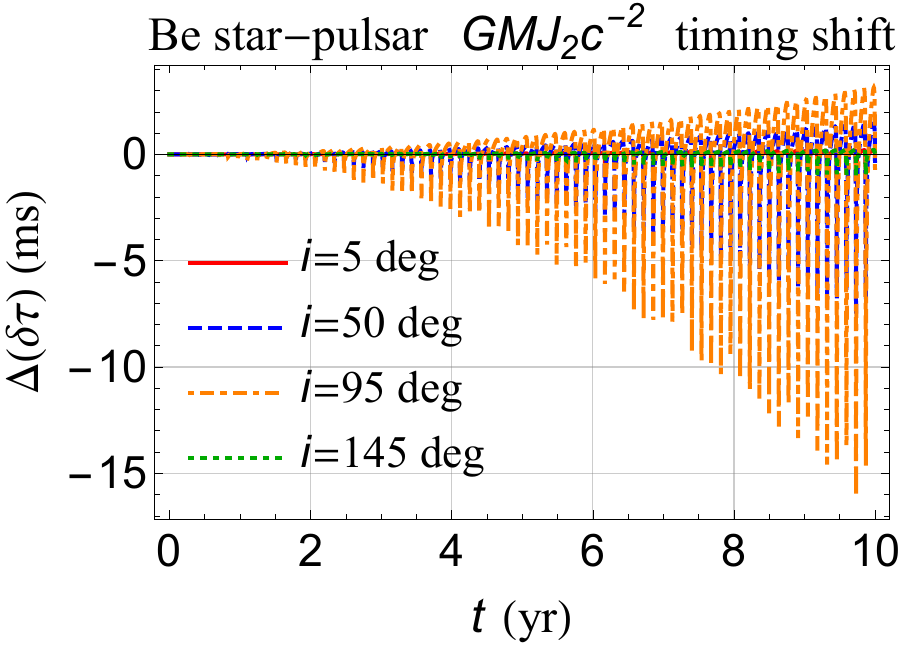}\\
\end{tabular}
}
}
\caption{Numerically integrated time series of the timing shift $\Delta\ton{\delta\uptau}$ due to \rfr{1pNMJ2}, in $\textrm{ms}$, for variations of the parameter space of a fictitious binary pulsar, characterized by
$\Pb = 50~\textrm{d},~e=0.8,~I=50~\textrm{deg},~\Omega=140~\textrm{deg},~\omega=149~\textrm{deg},~f_0=228~\textrm{deg}$, orbiting a Be-type star with $M = 15~\textrm{M}_\odot,~R_\textrm{e} = 5.96~\textrm{R}_\odot,~J_2 = 1.92\times 10^{-3},~i=60~\textrm{deg},~\phi=217~\textrm{deg}$.}\label{fig_1pNMJ2}
\end{center}
\end{figure}

\begin{figure}[ht]
\begin{center}
\centerline{
\vbox{
\begin{tabular}{cc}
\epsfysize= 4.4 cm\epsfbox{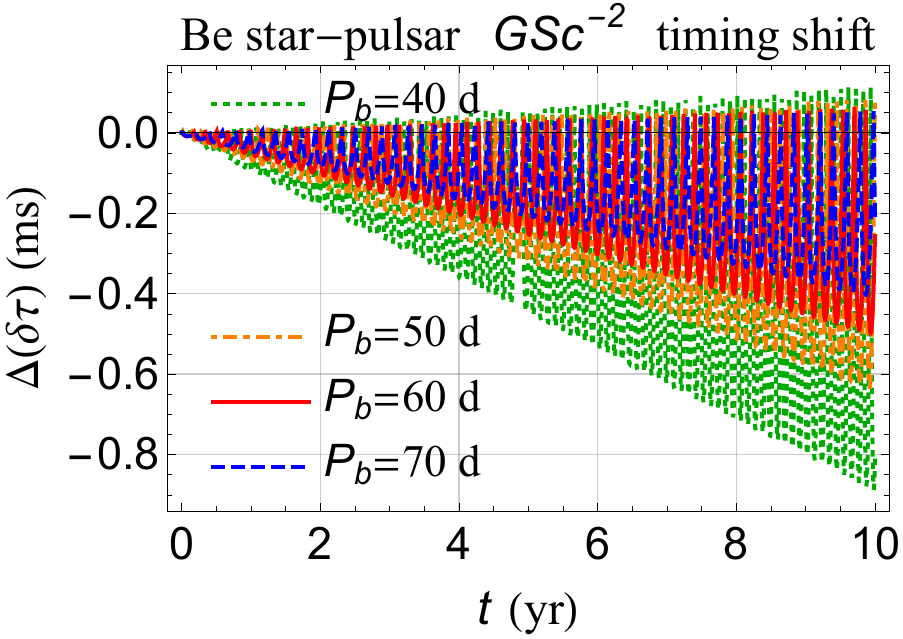}&\epsfysize= 4.4 cm\epsfbox{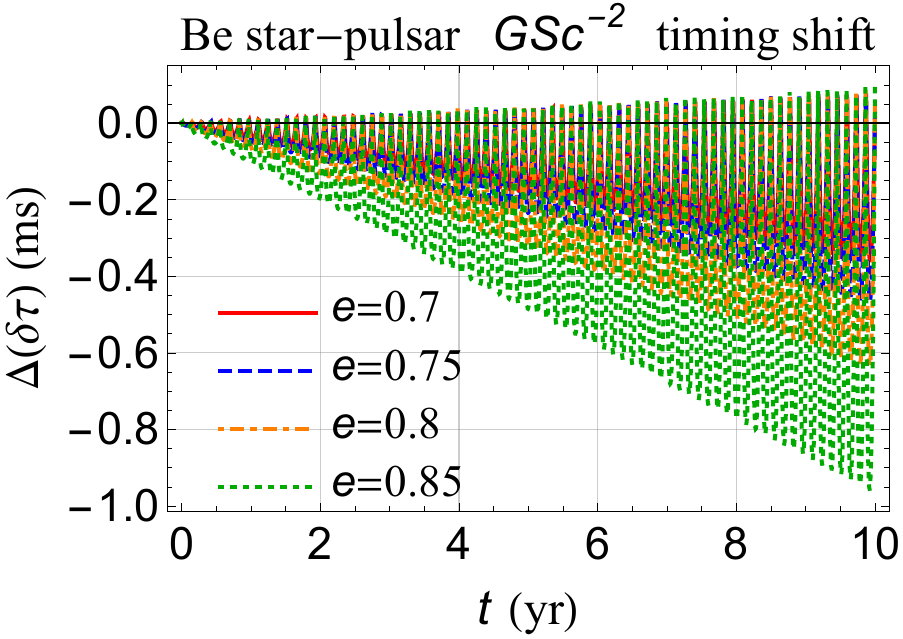}\\
\epsfysize= 4.4 cm\epsfbox{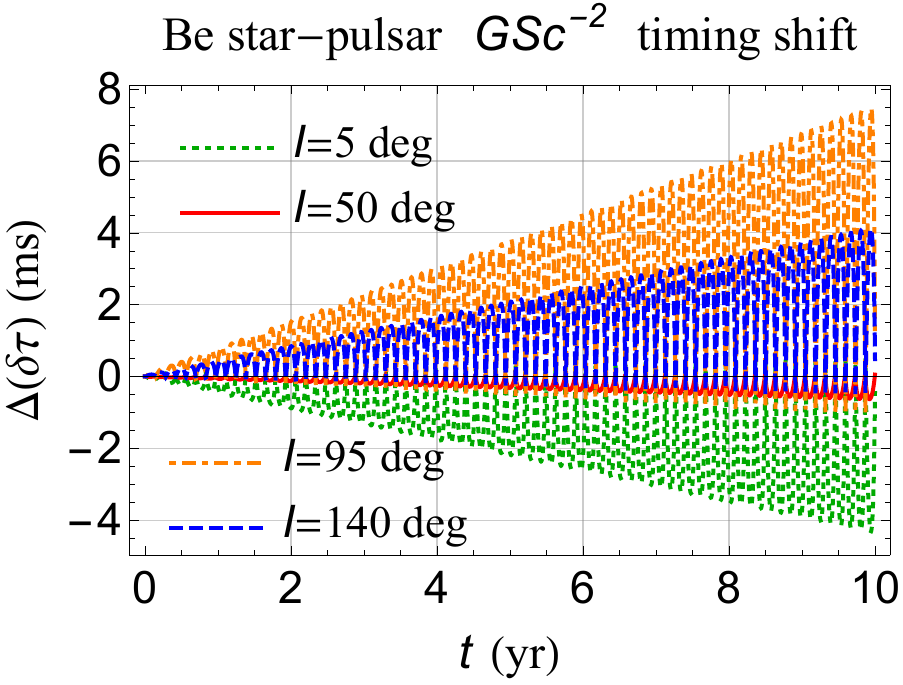}&\epsfysize= 4.4 cm\epsfbox{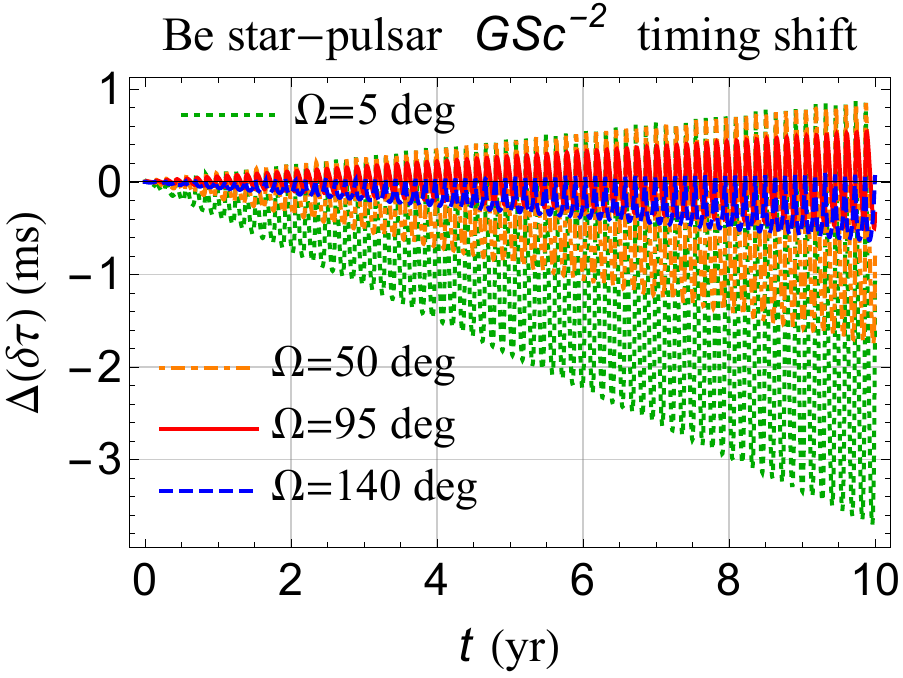}\\
\epsfysize= 4.4 cm\epsfbox{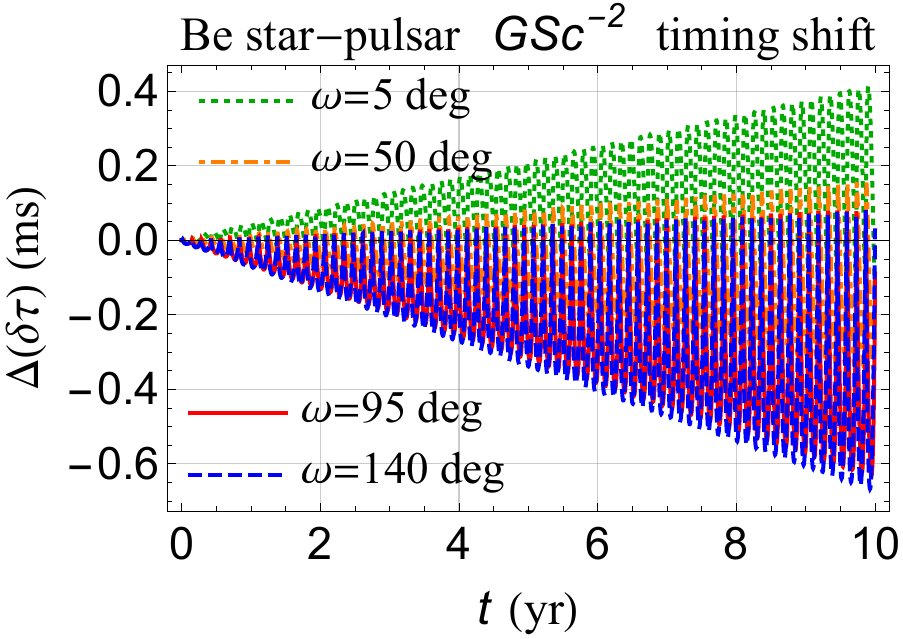}&\epsfysize= 4.4 cm\epsfbox{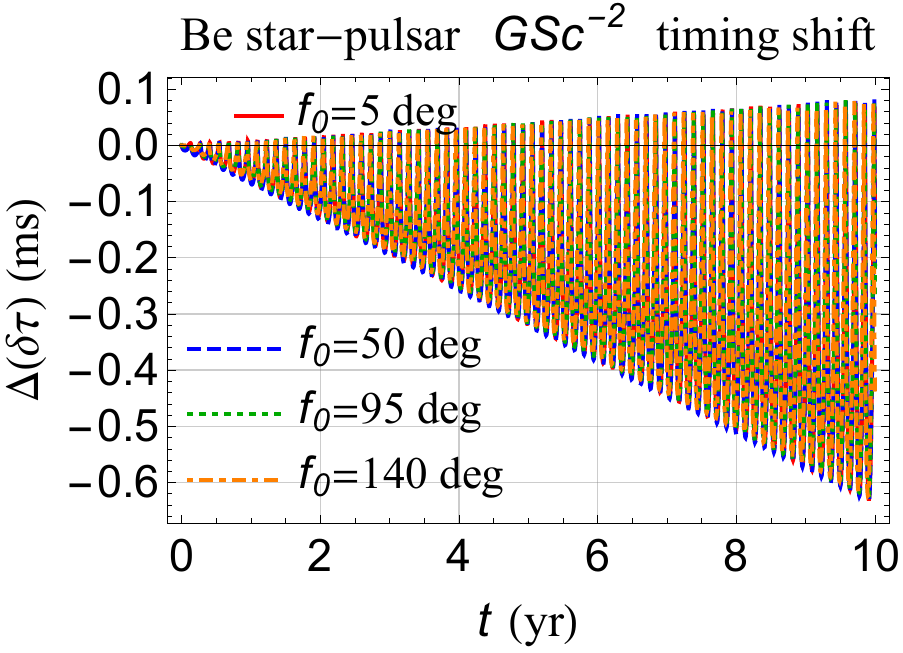}\\
\epsfysize= 4.4 cm\epsfbox{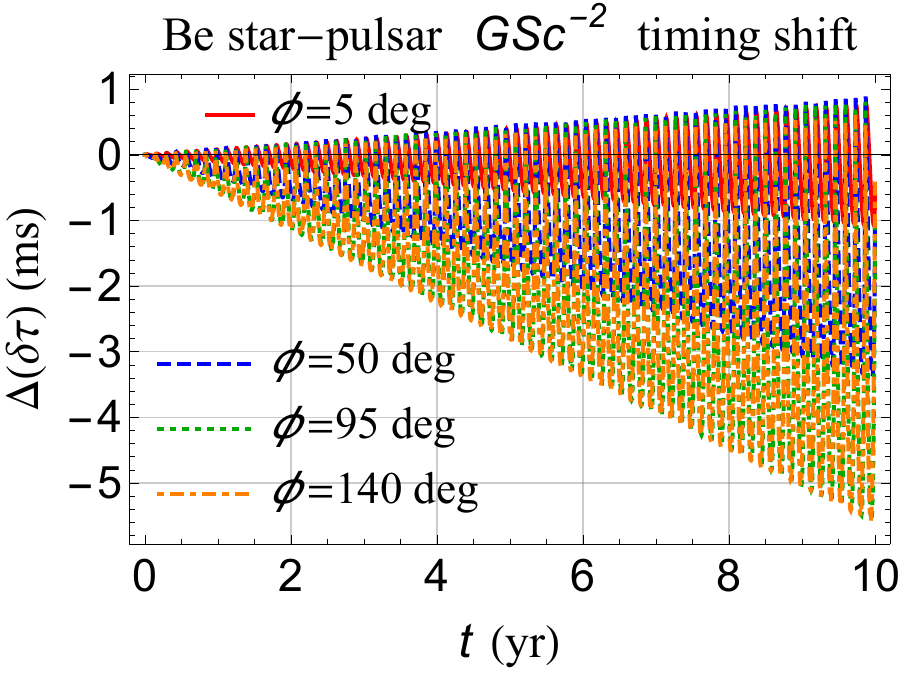}&\epsfysize= 4.4 cm\epsfbox{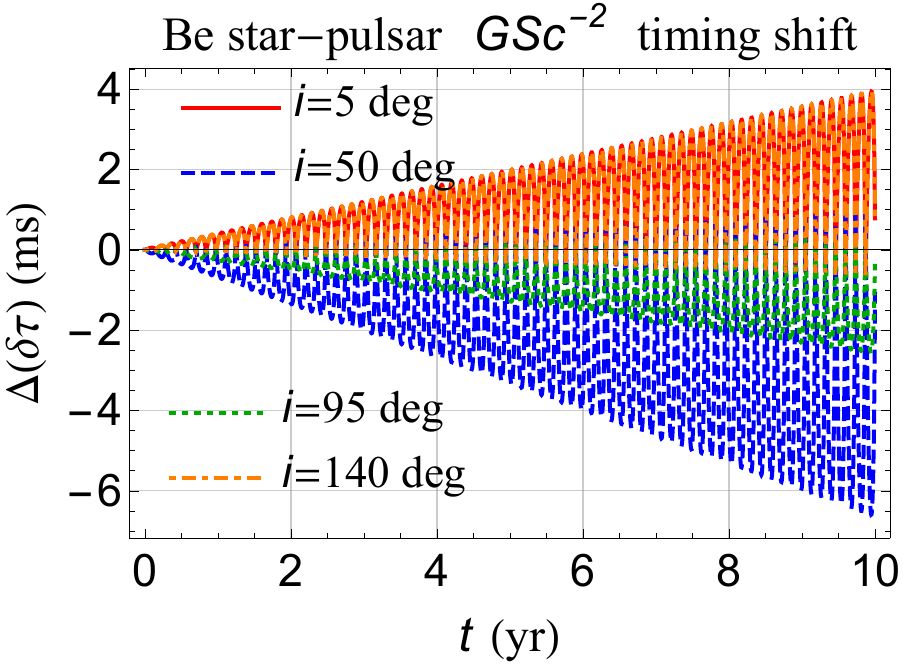}\\
\end{tabular}
}
}
\caption{Numerically integrated time series of the timing shift $\Delta\ton{\delta\uptau}$ due to \rfr{1pNS}, in $\textrm{ms}$, for variations of the parameter space of a fictitious binary pulsar, characterized by
$\Pb = 50~\textrm{d},~e=0.8,~I=50~\textrm{deg},~\Omega=140~\textrm{deg},~\omega=149~\textrm{deg},~f_0=228~\textrm{deg}$, orbiting a Be-type star with $M = 15~\textrm{M}_\odot,~S = 3.41\times 10^{45}~\textrm{J}~\textrm{s},~i=60~\textrm{deg},~\phi=217~\textrm{deg}$.}\label{fig_1pNS}
\end{center}
\end{figure}

\begin{figure}[ht]
\begin{center}
\centerline{
\vbox{
\begin{tabular}{cc}
\epsfysize= 4.4 cm\epsfbox{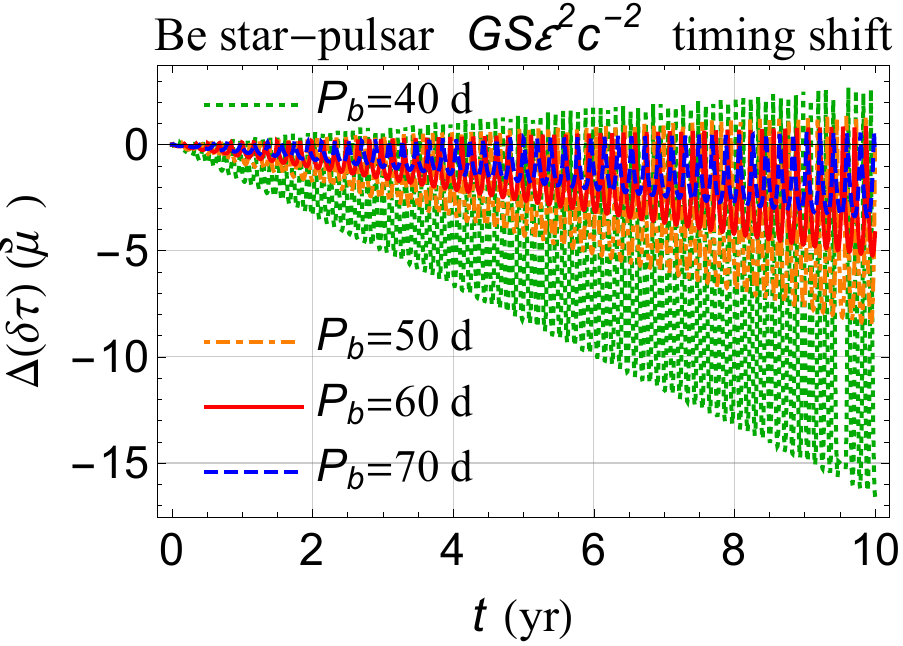}&\epsfysize= 4.4 cm\epsfbox{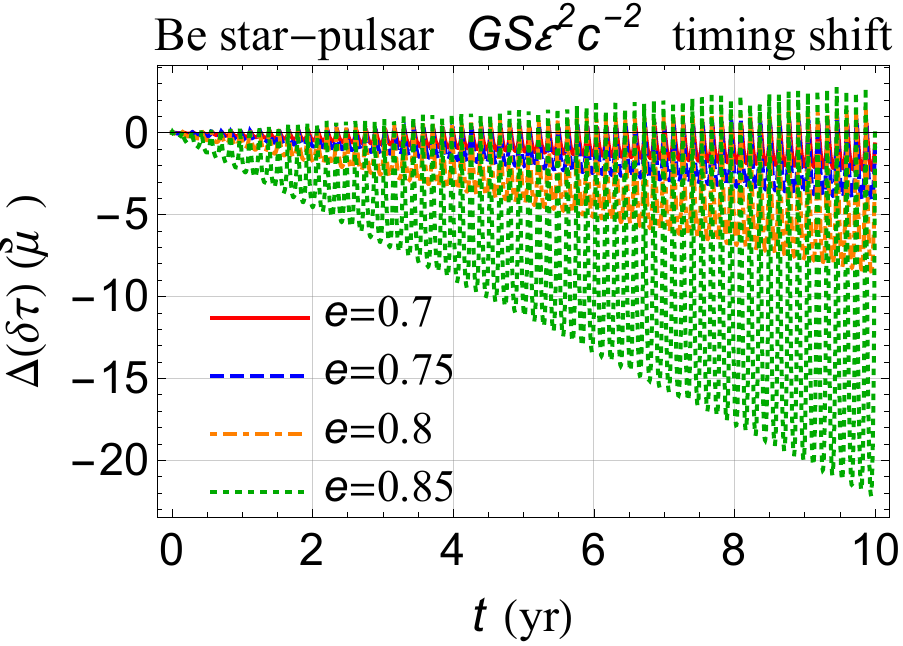}\\
\epsfysize= 4.4 cm\epsfbox{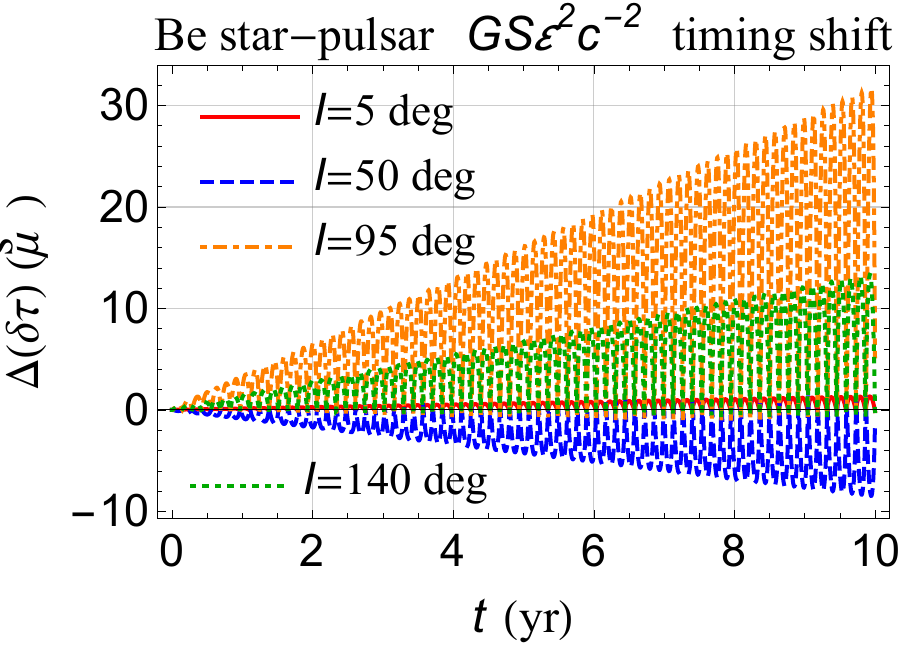}&\epsfysize= 4.4 cm\epsfbox{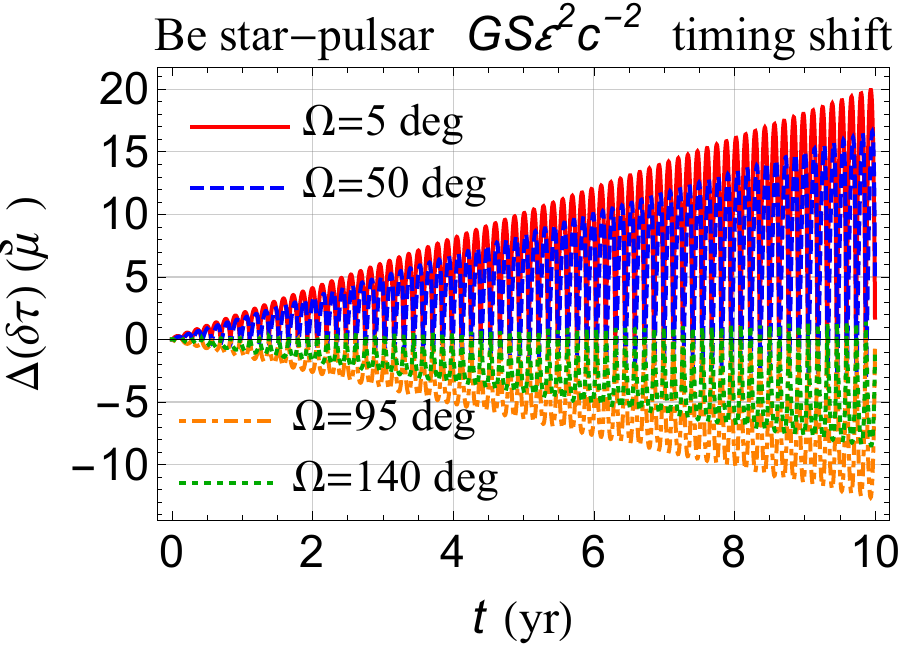}\\
\epsfysize= 4.4 cm\epsfbox{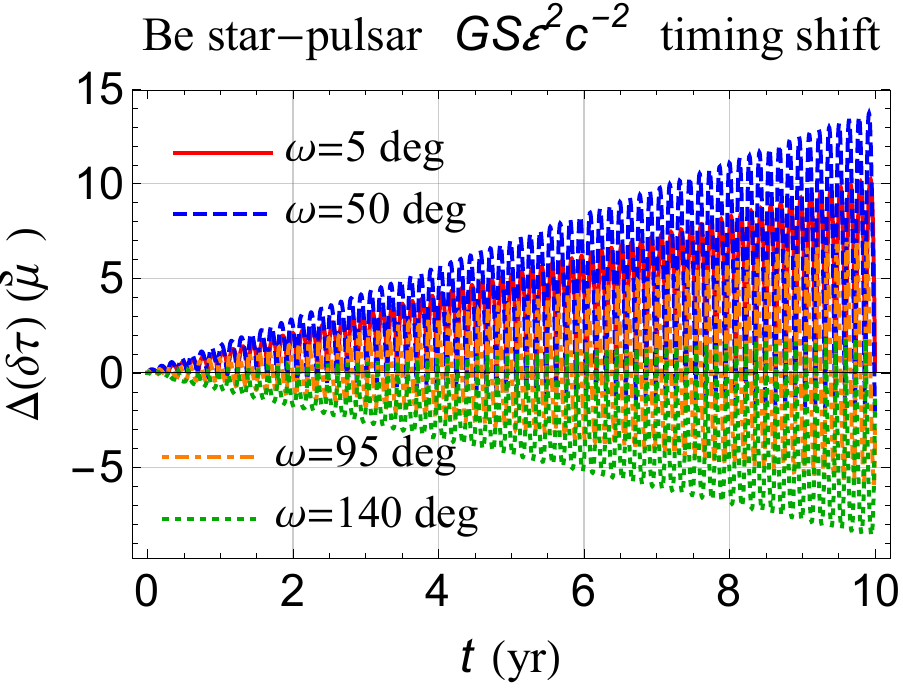}&\epsfysize= 4.4 cm\epsfbox{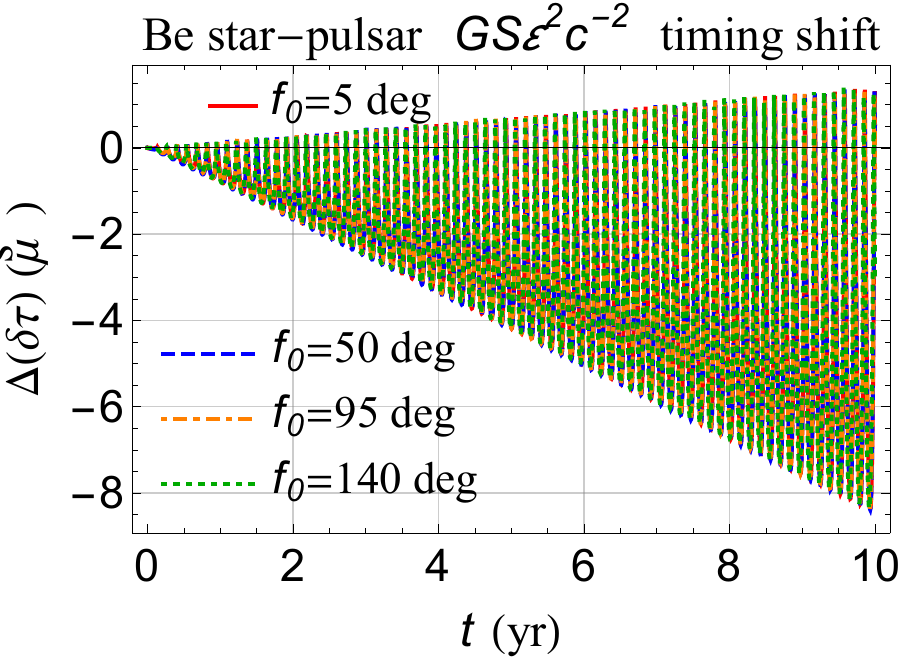}\\
\epsfysize= 4.4 cm\epsfbox{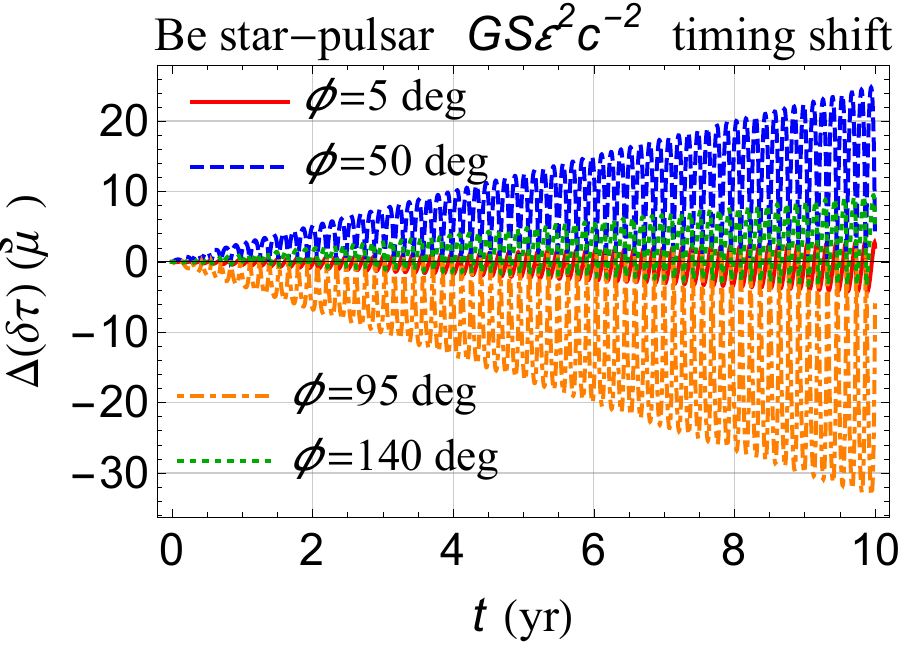}&\epsfysize= 4.4 cm\epsfbox{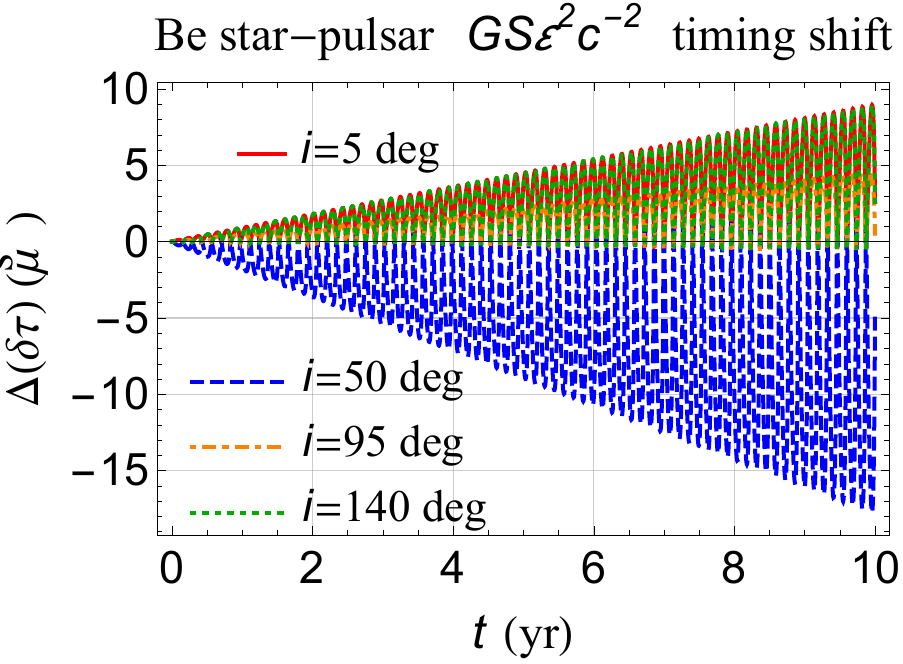}\\
\end{tabular}
}
}
\caption{Numerically integrated time series of the timing shift $\Delta\ton{\delta\uptau}$ due to \rfr{1pNSJ2}, in $\upmu\textrm{s}$, for variations of the parameter space of a fictitious binary pulsar, characterized by
$\Pb = 50~\textrm{d},~e=0.8,~I=50~\textrm{deg},~\Omega=140~\textrm{deg},~\omega=149~\textrm{deg},~f_0=228~\textrm{deg}$, orbiting a Be-type star with $M = 15~\textrm{M}_\odot,~R_\textrm{e} = 5.96~\textrm{R}_\odot,~\nu=0.203,~S = 3.41\times 10^{45}~\textrm{J}~\textrm{s},~i=60~\textrm{deg},~\phi=217~\textrm{deg}$.}\label{fig_1pNSJ2}
\end{center}
\end{figure}

The pK accelerations of \rfrs{NJ2}{1pNSJ2} perturb the otherwise Keplerian motion of the binary causing a change $\Delta\ton{\delta\tau}$ of the regular variation $\delta\tau$ of the TOAs due to the relative orbital motion of the pulsar and the massive companion. It can be modeled as the
ratio of the projection of the barycentric orbit of the pulsar onto the line of sight to $c$ \citep{1991PhRvL..66.2549D,2000ApJ...544..921K}. Thus, $\Delta\ton{\delta\tau}$ can be calculated by looking at the perturbations $\Delta z$ induced by \rfrs{NJ2}{1pNSJ2} on the $z$-component of the pulsar's barycentric orbital motion. We do that by numerically integrating the equations of motion of a fictitious pulsar \textcolor{black}{with mass $M_\mathrm{p}=1.4\,\mathrm{M}_\odot$} having as a companion a Be-type star with $M = 15~\textrm{M}_\odot,~R_\textrm{e} = 5.96~\textrm{R}_\odot,~\nu=0.203,~S = 3.41\times 10^{45}~\textrm{J}~\textrm{s},\,J_2 = 1.92\times 10^{-3}$, as per Table~\ref{Ester_zams}, for different values of its orbital configuration, determined by the initial values of the semimajor axis $a$, the eccentricity $e$, the orbital inclination $I$ to the plane of the sky, the longitude of the ascending node $\Omega$, the argument of periastron $\omega$, the true anomaly \textcolor{black}{at epoch} $f_0$, and of the stellar spin axis characterized by its inclination $i$ to the line of sight, and the longitude $\phi$ of the projection of the stellar spin  onto the plane of the sky. For a chosen pK acceleration ${\bds A}^\textrm{pK}$, in order to compute its perturbation $\Delta\ton{\delta\tau}=\Delta z\ton{t}/c$ over, say, 10 yrs, we perform two runs sharing the same initial conditions with and without ${\bds A}^\textrm{pK}$, calculate the resulting time series of $z\ton{t}/c$ and take their difference. Figures~\ref{fig_NJ2}~to~\ref{fig_1pNSJ2} depict our results for a given orbital configuration and the aforementioned Be-type main sequence star; in the panels of each Figure, we vary the Keplerian orbital elements and the orientation of $\kap$ in order to investigate the sensitivity to the parameter space of the adopted binary system. \textcolor{black}{For the sake of a comparison, the Keplerian delay for the adopted reference orbital configuration lies in the range $-250\,\mathrm{s}\lesssim\delta\tau\lesssim 50\,\mathrm{s}$ over one orbital revolution.}

Figure\,\ref{fig_NJ2} displays the Newtonian signatures due to the star's $J_2$. It can be noted that the resulting signal is strongly dependent on the initial value of the true anomaly, spanning the range from  $50\,\textrm{s}$ to $-150\,\textrm{s}$. Instead, for the given value $f_0 = 228\,\textrm{deg}$, the sensitivity to the other orbital parameters is rather modest, amounting to about $10\,\textrm{s}$.

The 1pN gravitoelectric Schwarzschild-like signatures due to the stellar mass monopole are reproduced in Figure\,\ref{fig_1pNM}.
Also in this case, the initial value of the true anomaly induces a marked variability in the decadal time series which ranges from to $-5\,\textrm{s}$ to $25\,\textrm{s}$. The other orbital parameters have less impact since the resulting variation of the signals is of the order of about $2\,\textrm{s}$.

The 1pN gravitoelectric time series induced by the quadrupole mass moment $J_2$ of the star are the subject of Figure\,\ref{fig_1pNMJ2}.
In this case, the ranges of variation due to all the orbital parameters are rather similar, amounting to about $20-50\,\textrm{ms}$.

Figure\,\ref{fig_1pNS} shows the 1pN gravitomagnetic Lense-Thirring signatures due to the spin dipole moment of the star. They are mainly sensitive to the orbital inclination $I$ and to the spin's inclination $i$ to the line of sight which induces a variability as large as $10-12\,\textrm{ms}$. The ranges of variation induced by the other parameters are, instead, of the order of $1-5\,\textrm{ms}$.

The 1pN gravitomagnetic time series caused by the spin octupole moment of the star are depicted in Figure\,\ref{fig_1pNSJ2}.
The widest range of variability, $30-50\,\upmu\textrm{s}$, is due to the inclination $I$, the node $\Omega$, and the longitude $\phi$ of the spin's projection onto the plane of the sky. We also checked that, for a pulsar orbiting in $\simeq 20-30\,\textrm{d}$ a star as massive as those in the last lines of Tables\,\ref{Ester_zams}\,to\,\ref{Ester_ev} with periastron distances of $r_\textrm{min}\simeq 1.2-1.1\,R_\textrm{eq}$, the magnitude of the timing signatures would reach the $\simeq 1-10\,\textrm{ms}$ level.
\subsection{The effects of the mass and spin multipoles of the pulsar}\lb{esteso}
Until now, we have considered the neutron star as a structureless, point particle moving around its more massive companion.
In fact, a pulsar is an extended body with its own mass and spin multipole moments which, at least in principle, may have an impact on its orbital motion in a full two-body framework.

The  modification of \rfr{1pNM} for two finite bodies of masses $M_\textrm{A},\,M_\textrm{B}$
is \citep{Sof89}
\begin{align}
{\bds A}^{\textrm{1pN}M} = \frac{\mu_\textrm{tot}}{c^2 r^2}\grf{\qua{\ton{4+2\zeta}\frac{\mu_\textrm{tot}}{r}-\ton{1+3\zeta}{\mathrm{v}}^2 +\frac{3}{2}\zeta{\mathrm{v}}_r^2}\bds{\hat{r}} + \ton{4-2\zeta}{\mathrm{v}}_r\textcolor{black}{\bds{v}}},\lb{megapN}
\end{align}
where $\mu_\textrm{tot}=GM_\textrm{tot}=G\ton{M_\textrm{A}+M_\textrm{B}}$, and $\zeta\doteq M_\textrm{A}M_\textrm{B}/M_\textrm{tot}^2$.
By taking the standard value $M_\textrm{p} = 1.4\,\textrm{M}_\odot$ for the mass of the pulsar, it is $\zeta = 0.078$ for the Be-star assumed in Section\,\ref{punto}. It turns out that the introduction of $\zeta$ in our numerical code changes the size of the time series of Figure\,\ref{fig_1pNM} by  $\simeq 0.1-0.2\,\textrm{s}$, while  their temporal patterns remain essentially unchanged. Given the current level of accuracy in the timing residuals, such a discrepancy might be significative, and \rfr{megapN} should be used instead of \rfr{1pNM}.

In regard to the angular momentum, the spin $\bds S$ of the Be-star in \rfr{1pNS}  should be replaced by the sum \citep{1975PhRvD..12..329B}
\eqi
\bds S\rightarrow \ton{1 + \frac{3}{4}\frac{M_\textrm{p}}{M}}\bds S +\ton{1 + \frac{3}{4}\frac{M}{M_\textrm{p}}}{\bds S}_\textrm{p},\lb{pulsarS}
\eqf
where ${\bds S}_\textrm{p}$ is the angular momentum of the pulsar.
By assuming, e.g., $S_\textrm{p}= 3\times 10^{40}\,\textrm{J}\,\textrm{s}$ as for PSR J0737-3039A \citep{2003Natur.426..531B,2004Sci...303.1153L,Kramer2012,
Kehletal017,2018IAUS..337..128K}, it turns out that the magnitude of the second term in \rfr{pulsarS} amounts to about $\simeq 7\times 10^{-5}$ of that of the first one. Moreover, it is  $\ton{3/4}\ton{M_\textrm{p}/M}=0.07$. Thus, as far as the 1pN gravitomagnetic Lense-Thirring effect is concerned, our scenario can be well approximated by a restricted two-body system with a spinning primary, and  \rfr{1pNS} is substantially adequate. Indeed, it turns out that rescaling the star's angular momentum in our numerical simulations as dictated by \rfr{pulsarS} slightly modifies the size of the time series of Figure\,\ref{fig_1pNS} by just $\simeq 1-1.5\,\textrm{ms}$.

The effect of the pulsar's quadrupole mass moment $J_2^\textrm{p}$  can be accounted for by the replacement $M\rightarrow M_\textrm{tot}$ in \rfr{NJ2} and by writing in its right-hand-side another term for $J_2^\textrm{p}$ analogous to that for the stellar oblateness $J_2$. As a result, by introducing the dimensional quadrupole mass moment
$Q_2 \doteq -J_2 M R_\textrm{e}^2$, the two terms in the right-hand-side of the resulting modified version of \rfr{NJ2} are weigthed by \citep{1975PhRvD..12..329B}
\begin{align}
\mathcal{Q}_2 \lb{Qu2} & = \ton{1+\frac{M_\textrm{p}}{M}}Q_2, \\ \nonumber \\
\mathcal{Q}_2^\textrm{p} & = \ton{1+\frac{M}{M_\textrm{p}}}Q^\textrm{p}_2.
\end{align}
For a neutron star, it is \citep{1999ApJ...512..282L,2004MNRAS.350.1416B,2012PhRvL.108w1104P,2013ApJ...777...68B}
\eqi
Q_2^\textrm{p} =  -q\frac{M_\textrm{p}^3 G^2}{c^4},
\eqf
with $0.07\lesssim q \lesssim 3.507$ for a variety of Equations of State (EOSs).
Thus, we have
\begin{align}
Q_2 & = -9.8\times 10^{47}\,\textrm{kg}\,\textrm{m}^2, \\ \nonumber \\
Q_2^\textrm{p} & = -q\,1.2\times 10^{37}\,\textrm{kg}\,\textrm{m}^2,
\end{align}
so that $\mathcal{Q}_2^\textrm{p}\simeq 10^{-10}\,\mathcal{Q}_2$. On the other hand, $M_\textrm{p}/M = 0.09$ in \rfr{Qu2}.
Thus, for the Newtonian signature of the quadrupole mass moment, the restricted two-body scenario with an oblate primary is adequate in the present case, and the use of \rfr{NJ2} is justified provided that the stellar quadrupole moment $Q_2$ is replaced by \rfr{Qu2}. Indeed, it turns out that the introduction of $\mathcal{Q}_2$ in the numerical integration changes the size of the time series in Figure\,\ref{fig_NJ2} by $\simeq 1\,\textrm{s}$ which may not be neglected, given the current level in $\upsigma_{\tau}$.

Despite \rfr{1pNMJ2} and \rfr{1pNSJ2} were derived  so far only for the motion of a test particle around a spinning, oblate mass, there are no doubts that they are adequate to the scenario considered here.

Actually, general relativity predicts that, in general, there is also a self-force due to the spin angular momentum of an extended rotating body in motion in an external gravitational field  which acts on it modifying its orbit through a spin-orbit coupling \citep{1951RSPSA.209..248P,1974RSPTA.277..59D,1979GReGr..11..149B,2006PhRvD..74l4006M,Mathis010,2012GReGr..44..719I}.
In order to quickly evaluate the possible impact of such an effect in our case, let us proceed as follows. To the 1pN level, the precession $\dot\Omega_{S_\textrm{p}}$ of, say, the node $\Omega$, averaged over one orbital revolution of the spinning pulsar  in its motion around the massive Be-type star, assumed nonrotating, is \citep{2012GReGr..44..719I}
\eqi
\dot\Omega_{S_\textrm{p}} = \frac{3\mu \sigma_\textrm{p}\csc I\ton{{\bds{\hat{\sigma}}}_\textrm{p}\bds\cdot\bds{\hat{m}}}}{2c^2a^3\ton{1-e^2}^{3/2}}.\lb{spinor}
\eqf
In this expression, ${\bds\sigma}_\textrm{p} = {\bds S}_\textrm{p}/M_\textrm{p}$ is the pulsar's spin angular momentum per unit mass, while $\bds{\hat{m}} =\grf{-\cos I\sin\Omega,\,\cos I\cos\Omega,\,\sin I}$ is a unit vector in the orbital plane perpendicular to the line of the nodes.
Let us compare \rfr{spinor}  to  the analogous precession induced by some of the effects previously considered in which the pulsar was treated as a point particle.  The Lense-Thirring node rate, induced by \rfr{1pNS} which is responsible for the $\simeq\textrm{ms}$ time series of Figure\,\ref{fig_1pNS},
is \citep{2012GReGr..44..719I}
\eqi
\dot\Omega_\textrm{LT} = \frac{2GS\csc I
\ton{\bds{\hat{S}}\bds\cdot\bds{\hat{m}}}}{c^2 a^3\ton{1-e^2}^{3/2}}.
\eqf
It turns out that, in the case of a pulsar like, e.g., PSR J0737-3039A orbiting the Be-type star of the Figures\,\ref{fig_NJ2}\,to\,\ref{fig_1pNSJ2}, it is $\left|\dot\Omega_{S_\textrm{p}}/\dot\Omega_\textrm{LT}\right|\simeq 10^{-5}$. Thus, we conclude that the spin-orbit self-force experienced by the pulsar is completely negligible in our scenario.
In fact, there is also a further self-force acting on an extended rotating body moving in an external gravitational field due to the spin-spin coupling between the angular momenta of the source and of the orbiter itself \citep{1979GReGr..11..149B,2012GReGr..44..719I}. By following the same strategy for the spin-orbit coupling, the results in \citet{2012GReGr..44..719I} allow to conclude that, in our case, such an effect is even smaller than the previous one, being of the order of $\simeq 10^{-7}$ of, say, the $\simeq \textrm{ms}$-level Lense-Thirring signal.

\section{Summary and conclusions}\lb{conclu}
We preliminarily  explored the possibility of putting to the test several pK features of motion of Newtonian and pN origin in binaries hosting a pulsar and a massive, fast rotating, highly distorted main sequence star characterized by mass $M$, angular momentum $\bds S$, equatorial and polar radii $R_\textrm{e},~R_\textrm{p}$, flattening $\nu$, ellipticity $\varepsilon$ and dimensionless quadrupole moment $J_2$. Indeed, in addition to the usual 1pN Schwarzschild and Lense-Thirring effects due to the mass monopole $\ton{\propto GM c^{-2}}$ and spin dipole $\ton{\propto GSc^{-2}}$ moments, respectively, of the distorted stellar field, there are also other 1pN orbital effects, induced by the mass quadrupole $\ton{\propto GMR^2_\textrm{e}J_2 c^{-2}}$ and spin octupole $\ton{\propto GSR^2_\textrm{e}\varepsilon^2c^{-2}}$ moments, whose magnitudes may, perhaps, lie above the sensitivity threshold of the pulsar timing residuals in yet-to-be-discovered close binaries. However, the Newtonian perturbations due to $J_2$ are larger than the pN ones.

In order to perform a preliminary sensitivity analysis, we numerically investigated the orbital shifts $\Delta\ton{\delta\tau}$ induced over 10 yr by all of such pK perturbations on the otherwise Keplerian R{\o}mer-type delay $\delta\tau$ in the pulsar's TOAs for a Be-type main sequence star characterized by  $M = 15~\textrm{M}_\odot,~R_\textrm{e} = 5.96~\textrm{R}_\odot,~\nu=0.203,~S = 3.41\times 10^{45}~\textrm{J}~\textrm{s},\,J_2 = 1.92\times 10^{-3}$ orbited by a pulsar with an orbital geometry compatible $\ton{\Pb\simeq 40-70~\textrm{d}}$ with some of the tightest binaries of this kind out of those discovered so far. We also investigated the sensitivity of the pK timing shifts to the whole system's parameter space by varying both the orientation of the stellar spin axis and the orbital elements of the pulsar's orbit.
It turns out that the magnitude of the Newtonian signature due to $J_2$ can be as large as $\lesssim 4-150~\textrm{s}$,
while the pN gravitoelectric quadrupolar one is $\lesssim 10-40~\textrm{ms}$.
The pN gravitoelectric (Schwarzschild-like) signal due to the stellar mass monopole can be as large as $\lesssim 1.5-20~\textrm{s}$ level.
The pN gravitomagnetic shifts due to the spin dipole (Lense-Thirring) and octupole moments, for the evaluation of whose size the knowledge of the stellar spin angular momentum $S$ and ellipticity $\varepsilon$ is crucial, are of the order of $\lesssim 0.5-6~\textrm{ms}$, and $\lesssim 5-20~\upmu\textrm{s}$, respectively.
The rms of the timing residuals of all the non-recycled, non-millisecond pulsars like those having a fast rotating main sequence star companion discovered so far are $\lesssim \textrm{ms}$ over $2-13~\textrm{yr}$. It seems difficult that they can be substantially improved in the future. This implies that, in principle, all the pK effects considered fall within the potential measurability domain, except the pN  spin octupole which is $\simeq 1-2$ orders of magnitude weaker, at least for the orbital configurations and the star considered in this study. The pN Lense-Thirring signatures are just at the $\simeq\textrm{ms}$ level.
The different temporal patterns characteristic of the signals investigated may be helpful in separating them. It turns out that the tiniest pN effect may reach the $\simeq 1-10\,\textrm{ms}$ level only for a very tight, eccentric binary $(P_\textrm{b}\simeq 20-30\,\textrm{d},\,r_\textrm{min}\simeq 1.1-1.2\,R_\textrm{e})$ hosting a Be-star with $30\,\textrm{M}_\odot,\,S=125-870\times 10^{44}\,\textrm{J\,s}$.
%, apart from  PSR J1903+0327 whose residuals are at the $\simeq %\upmu\textrm{s}$ level over $\simeq 3~\textrm{yr}$.

Finally, we stress the preliminary nature of our sensitivity analysis. To this aim, we remark that we did not compute the other kinds of time delay connected, e.g., with the propagation of the electromagnetic waves in the deformed spacetime.  Moreover, we did not perform a full covariance analysis implying a simulation of the pulsar's TOAs, their reduction, and parameter estimation.
\section*{Acknowledgements}
We are grateful to A. Possenti for useful information. MR and ADS also acknowledge the strong support of the French Agence Nationale de la Recherche (ANR), under grant ESRR (ANR-16-CE31-0007-01).
%-----------------------------------------
%
%
%
%
%
%
%
%
%
%
%
%
\bibliography{MR_biblio,MS_binary_pulsar_bib,Gclockbib,semimabib,PXbib}{}

\end{document}